\documentclass[a4paper,11pt]{article}
\usepackage{jheppub} 
\usepackage{amsmath}
\usepackage{amssymb}
\usepackage{amsfonts}
\usepackage{amsthm}
\usepackage{amsbsy}
\usepackage{array}
\usepackage{mathrsfs}
\usepackage{physics}
\usepackage{mathtools}
\usepackage{hyperref}
\usepackage{subcaption}
\usepackage{soul}
\usepackage{dsdshorthand}

\usepackage{tikz}
\usetikzlibrary{decorations.markings}

\title{\boldmath Asymptotic charges as detectors and the memory effect in massive QED and perturbative quantum gravity}

\author[a]{Ian Moult,}
\author[a]{Brett Oertel,}
\author[b]{Sabrina Pasterski}
\affiliation[a]{Department of Physics, Yale University,New Haven, CT 06511}
\affiliation[b]{Perimeter Institute,Waterloo, ON N2L 2Y5, Canada}

\emailAdd{ian.moult@yale.edu}
\emailAdd{brett.oertel@yale.edu}
\emailAdd{spasterski@perimeterinstitute.ca}

\abstract{It has been shown that there are an infinite set of asymptotic symmetries in quantum gravity and QED, and this has been extended to dressed states in some cases. Here we rederive these statements in terms of detectors in order to clarify, confirm, and generalize these results to include external hard gravitons. Using detectors and including the full $t$ dependence in Faddeev-Kulish dressings allows us to correct discrepancies in the literature and make new statements. We show that Faddeev-Kulish dressings correctly encode the memory effect in the `in' and `out' scattering Fock spaces. We find a physical contribution to the memory eigenvalues arising from the dressings in both cases. 
 }

\begin{document}
\definecolor{yaleblue}{rgb}{0.06, 0.3, 0.57}
\newcommand\BOnote[1]{{\color{yaleblue}{\bf BO: #1}}}
\maketitle
\flushbottom

\newpage 

\section{Introduction} 
\label{sec:intro}
It has been known for some time that one can construct infinitely many charges which are asymptotically conserved during scattering in a theory coupling matter to a massless spin 2 field and that conservation of these charges is equivalent to the soft graviton theorem \cite{Strominger_2014, He_2014, Campiglia_2015}. The same is true for spin 1 \cite{He_2015, Campiglia:2015qka}. The central aim of this work is to re-derive and better understand these results using detector (light-ray) operators. This further contributes to a number of interesting recent relations between the celestial holography program and light-ray operators \cite{Hu:2023geb,Gonzalez:2025ene,Moult:2025njc,Himwich:2025ekg,Sheta:2025oep,Strominger:2026yrh, Hirai_2019, Hirai_2021, Hirai_2023}.  Since detectors are well-understood objects which can be defined entirely within the bulk, we hope that this will serve to promote understanding of the aforementioned results for those unfamiliar with them, as well as provide a convenient mathematical framework for those already familiar with them. Furthermore, we find that working with detectors gives some results which correct the existing literature, see Section \ref{conclusion} for a list of these. We also discuss and physically interpret these conservation laws in the context of Faddeev-Kulish dressed scattering states, and gravitational and electromagnetic memory.

Since detectors are defined as (integrals of) operators inserted at the boundary of spacetime, they can be used to describe measurements performed asymptotically far away from a scattering process. Indeed, this is exactly what happens at particle colliders, and so it is not surprising that detectors have been studied since the early days of QCD \cite{PhysRevD.6.3543, BRANDT1971541, Sterman:1975xv, PhysRevD.19.2018}. More recently, a surge of interest in light-ray operators has occurred and resulted in detailed studies in the context of conformal field theories \cite{Hofman_2008, Kravchuk_2018, kologlu2020lightrayopeconformalcolliders}. For a detailed review, see \cite{Moult:2025nhu}. Detectors can be thought of as light-ray operators placed at infinity, and we now know how to perturbatively define and renormalize them in various theories, for example in Wilson-Fisher theory \cite{Caron_Huot_2023} and in the critical $O(N)$ model \cite{li2025reggetrajectoriesdetectorsdistributions}. As operators at the boundary of Minkowski space, detectors are naturally suited to describe the asymptotic charges discussed in this work, and since a wealth of mathematical knowledge about detectors already exists, doing so is a natural step towards improving understanding of these charges nature. Furthermore, due to the difficulty in defining local observables in quantum gravity, detectors are a natural choice for investigating perturbative quantum gravity. Recent work inspired by this idea can be found in \cite{herrmann2024energycorrelatorsperturbativequantum,Ruan:2026xyd,Gonzo:2020xza,Chicherin:2025keq}.

We begin by discussing detectors, focusing on their precise definition as distribution valued operators and the associated mathematical nuances. We will then move on to rigorously defining asymptotic charges in massive scalar QED and gravity coupled to a massive scalar and proving their conservation for arbitrary scattering states. We show conservation both for ordinary scattering states, and Faddeev-Kulish dressed scattering states. In doing so we find that certain commutators in the literature appear to be in tension with conservation for dressed states. By explicitly including the $t$ dependence in the dressing, we are able to extend this result in gravity to the general case of scattering of external hard gravitons and scalars, which to the best of our knowledge has not been done before. We find that in both QED and gravity Faddeev-Kulish states diagonalize the asymptotic charges and the memory operator, and we derive the contributions to the memory eigenvalue from the dressing. We find that, unlike previously stated in \cite{Choi_2018}, the dressing does contribute to the gravitational memory eigenvalues and the BMS charges are not zero for dressed states. 

The BMS charges we find are physical in origin and we show that they agree with the predictions of Thorne, Braginsky,  Bieri and Garfinkle \cite{1987Natur.327..123B, PhysRevD.45.520, Bieri_2013, Bieri_2014}.\footnote{From now on we will speak about matching to Bieri and Garfinkle's more recent papers, but we wish to emphasize that the gravitational results also appeared much earlier in the papers by Braginsky and Thorne.} In QED the BMS charge is the $\ell=0$ part of the memory due to the outgoing state, and in gravity it is the $\ell=0$ and $\ell=1$ parts. These cancel due to charge (QED) or energy and momentum (gravity) conservation when considering the full memory due to the outgoing and incoming state. However, since these terms result from considering the physical (gauge fixed) Fock space, they should still be included in the memory eigenvalues.

\vspace{1em}\noindent

\section{An introduction to detectors}
\label{sec2}
\subsection{Distribution valued operators}
Despite being successfully used by physicists for decades to make some of the most precise experimental predictions in science, most quantum field theories have thus far escaped precise mathematical formulation. A quantum field theorist must know when to ignore mathematical subtleties and when it is judicious to be more pedantic. Since some of the results in this paper rely on distributional analysis, we will take a small amount of time here to explain both the role of distribution theory in more rigorous attempts at quantum field theory, and why it is useful when working with detectors.

At least as far as scattering theory is concerned, one could claim to have a satisfactory mathematical formulation of a quantum field theory (at a particular order in perturbation theory) if one manages to precisely define the in and out asymptotic Fock spaces and the physical map between them known as the S-matrix. To be specific, one would need to show that these Fock spaces are separable Hilbert spaces which contain all possible required physical states, that the S-matrix is well-defined and unitary, and that the appropriate symmetries are obeyed (for a recent and more complete discussion on the requirements of such a construction for various theories see \cite{Prabhu:2022zcr} which is closely related to this work). It is well-known that a construction like this is not difficult to achieve for a free, scalar quantum field theory, one need only use the Wightman axioms \cite{Wightman1965FIELDSAO}. The key idea behind the Wightman axioms is to use distribution valued operators to deal with the divergences that appear in elementary quantum field theory calculations. For example, consider a free, scalar quantum field $\phi(x)$ with action 
\begin{equation}
    S = \int d^4x \frac{1}{2}\partial_{\mu}\phi(x)\partial^{\mu}\phi(x) \ .
\end{equation}
One immediately runs into a divergence if one naively tries to compute the time-ordered two-point function as
\begin{equation}
    \bra{0}T\{\phi(x)\phi(y)\}\ket{0}= \int \frac{d^3p}{(2\pi)^3}\frac{1}{2\w}e^{-i\w|x^0-y^0|+i\bp\cdot(\bx-\by)} \ .
\end{equation}
However, we should really consider this integral as a tempered distribution which acts on Schwartz functions of $x$ and $y$. Thus, we actually define the time-ordered two point function using a regularization procedure, so that it acts on Schwartz functions $f(x)\in \cS(\R^4)$ as follows:
\begin{equation}
    f\rightarrow \lim_{\e\rightarrow 0}\int d^4x\int \frac{d^3p}{(2\pi)^3}\frac{1}{2\w}e^{-i\w(|x^0-y^0|-i\e)+i\bp\cdot(\bx-\by)}f(x) \ ,
\end{equation}
which is finite. One of the claims of the Wightman axioms is that there exists a consistent set of regularization procedures for all possible correlation functions such that they are well-defined distributions over functions of the spacetime or momentum variables. For example, for a scalar theory in d dimensions, any operator 
\begin{equation}
    \cO(x_1,\cdots,x_i) \ ,
\end{equation}
should yield a well-defined distribution of Schwartz functions of the $x_i$ variables when inserted into a correlation function:
\begin{equation}
    \bigg|\int d^dx_1\cdots d^dx_i\bra{0}\cO(x_1,\cdots,x_i)\ket{0}f_1(x_1)\cdots f_i(x_i)\bigg|<\infty \ , \hspace{0.5cm} f_i\in \cS(\R^d) \ .
\end{equation}
This construction is possible for a free quantum field theory, but issues immediately arise when one includes interactions. The easiest to deal with are so-called UV divergences, which in the modern Wilsonian viewpoint occur due to unknown high energy physics and are dealt with using perturbative renormalization (of course we are assuming the interactions are weak). In fact, these divergences are not necessarily disastrous even from a formal point of view, as shown by Lehmann, Symanzik and Zimmerman, Haag and Ruelle, and others, in works which established that massive quantum field theories should admit well-defined scattering theory in spite of these UV divergences \cite{1955NCimS...1..205L, PhysRev.112.669, ruelle1962asymptotic}. In essence, one can technically deal with divergences through regularization, which is physically justified by viewing the theory as an effective field theory. This means UV divergences do not interfere with attempts to define the S-matrix at a given order in perturbation theory.

However, one also runs into IR divergences in interacting theories, and these are more subtle. For example, due to the soft theorem, we encounter a divergence in S-matrix elements as the energy of an external graviton or photon goes to zero. We will see that these divergences are closely related to the conservation laws that we will derive. Detectors also suffer from IR divergences and require renormalization in perturbation theory. How to carry this out in practice is discussed in detail in \cite{Caron_Huot_2023}. It important to view detectors as distribution valued operators in order to understand their divergences. For example, when calculating in-in event shapes of the detectors in the leading Regge trajectory of Wilson-Fisher theory, the authors of \cite{Caron_Huot_2023} obtain a function like
\begin{equation}
    \sim \theta(-p^2)(z\cdot p )^{J_L}(-p^2)^{\frac{d-4}{2}-J_L-2} \ ,
\end{equation}
where $z$ is null. Whilst this naively appears to be finite, if one integrates it against a test function in $p$ which is non-zero when $p=z$, one obtains a divergence in $d=4-\e$ of the form $\sim \frac{1}{\e}$. Thus, we see that it is crucial to consider detectors in the context of distribution valued operators in order to understand them. We will also see examples later in this work where this viewpoint is essential.
\subsection{Detectors}
Throughout this paper we will define detectors as we use them so that calculations are self-contained, but it will be useful to briefly discuss detectors in general here. In this discussion we choose to work with a real, massless scalar field $\phi(x)$ since it captures all important physical properties but also allows us to write down explicit formulas which are relatively simple. As a start, let's note that an essential physical property of detectors is that they describe measurements made asymptotically far away from a physical event. Thus, we will want to consider operators at future null infinity $\cI^+$. Note that null infinity is chosen since $\phi$ is massless, otherwise we would want to choose future timelike infinity. We do use some massive detectors later when working with massive scalar fields, but within this section we will only consider a massless field. To send $\phi$ to future null infinity we can consider
\begin{equation}
    \phi(u,y)\equiv\lim_{L\rightarrow\oo}L^{\Delta_{\phi}}\phi(x+Ly) \ . 
\end{equation}
where $u\equiv-x\cdot y$ , $y=(1, \hat{\by})$ is a future-pointing null vector, $\Delta_{\phi}=\frac{d-2}{2}$ , and the factor $L^{\Delta_{\phi}}$ is to cancel out the scaling of $\phi$ under dilatations. Together, $u$ and $y$ label where along $\cI^+$ we are sending $\phi$ to. In particular, $y$ specifies a direction on the celestial sphere, and $u$ specifies a retarded time. 

\begin{figure}[h!]
\centering
\hspace{3cm}
\begin{tikzpicture}

  \coordinate (itop)   at (0,4);    
  \coordinate (ibot)   at (0,-4);   
  \coordinate (ileft)  at (-4,0);   
  \coordinate (iright) at (4,0);    

  \begin{scope}
    \clip (itop) -- (iright) -- (ibot) -- (ileft) -- cycle;

    \fill[blue!15!white] (itop) -- (iright) -- (ibot) -- (ileft) -- cycle;

    \draw[blue!50!black, thick, 
          decoration={markings, mark=at position 0.7 with {\arrow{>}}},
          postaction={decorate}]
      (-3.5, -2.5) .. controls (-1.0, 0.0) and (0.0,1.0) .. (2.5, 3.5);

    \draw[blue!50!black, thick, 
          decoration={markings, mark=at position 0.7 with {\arrow{>}}},
          postaction={decorate}]
      (0, -4) .. controls (2.0, -1.5) and (-2, 1.5) .. (0, 4);

  \end{scope}

  \draw[blue!40!black, thick] (itop) -- (iright);
  \draw[blue!40!black, thick] (iright) -- (ibot);
  \draw[blue!40!black, thick] (ibot) -- (ileft);
  \draw[blue!40!black, thick] (ileft) -- (itop);

  \node[rotate=0, blue!50!black, font=\itshape] at (-2, 0.0) {light-ray};

  \node[rotate=0, blue!50!black, font=\itshape] at (1.5, -0.6) {time-like path};

  \node[rotate=-45, blue!40!black, font=\small] at (1.8, 1.8) {$r = +\infty$};

  \node[rotate=45, blue!40!black, font=\small] at (1.8, -1.8) {$r = -\infty$};


  \node[font=\large] at (0, 4.3) {$i^+$};

  \node[font=\large] at (0,-4.3) {$i^-$};

  \node[right=2pt, font=\large] at (4, 0.0) {$i^0$};


  \node[font=\small, anchor=west] at (0.5, 3.6) 
    {$\mathcal{I}^+_+ = S^2, \quad (u = +\infty)$};

  \node[font=\small, anchor=west] at (2.1, 2.1) 
    {$\mathcal{I}^+ = \mathbb{R} \times S^2$};

  \node[font=\small, anchor=west] at (3.6, 0.5) 
    {$\mathcal{I}^+_- = S^2, \quad (u = -\infty)$};

  \node[font=\small, anchor=west] at (2.1, -2.1) 
    {$\mathcal{I}^- = \mathbb{R} \times S^2$};

\end{tikzpicture}
\caption{A Penrose diagram of Minkowski spacetime. Future timelike infinity ($i^+$), past timelike infinity ($i^-$), spatial infinity ($i^0$), future null infinity ($\mathcal{I}^+$), and past null infinity ($\mathcal{I}^-$) are shown. If one takes the limit $u\rightarrow+\infty$ on $\mathcal{I}^+$ then one ends up at $\mathcal{I}^+_+$, and if one takes the limit $u\rightarrow-\infty$ on $\mathcal{I}^+$ then one ends up at $\mathcal{I}^+_-$. }
\end{figure}
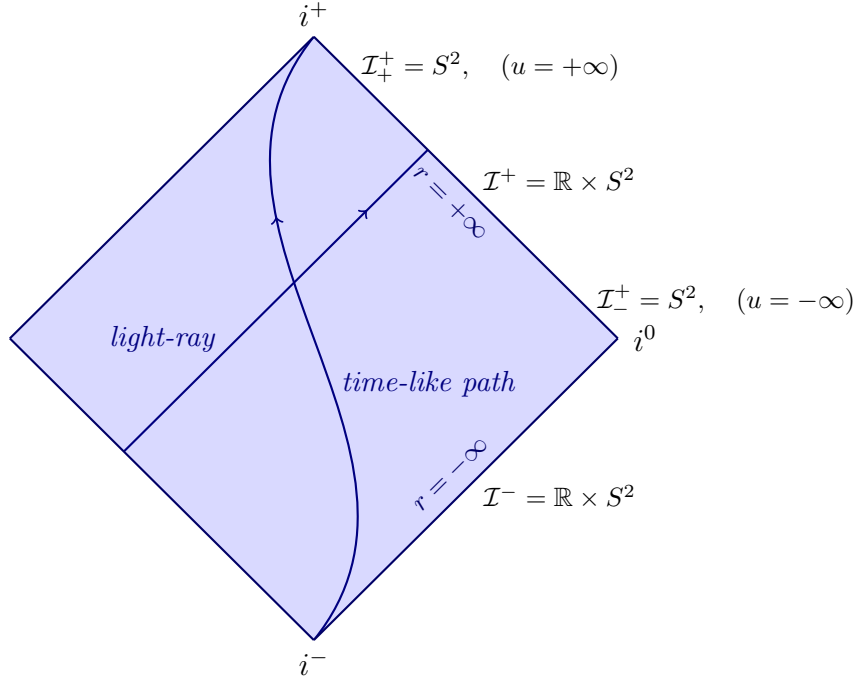
By expanding $\phi(u,y)$ as
\begin{equation}
    \lim_{L\rightarrow\oo}L^{\frac{d-2}{2}}\int \frac{d^{d-1}p}{(2\pi)^{d-1}}\frac{1}{2\w_p}(a(p)e^{ip(Lx+y)}+a^{\dagger}(p)e^{-ip(x+Ly)}) \ ,
\end{equation}
and performing a stationary phase approximation (we will discuss the legitimacy of this approximation soon) one can informally write
\begin{equation}
    \phi(u,y) =  \frac{i}{2(2\pi)^2}\int^{\oo}_{0}d\w(-a(\w \hat{y})e^{-i\w u}+a^{\dagger}(\w \hat{y})e^{i\w u}) \ .
\end{equation}
More generally, we can consider operators with spin, like a traceless, symmetric operator $\cO_{\mu_1...\mu_J}$. To send this to future, null infinity we could use
\begin{equation}
    \cO_{\mu_1\cdots\mu_J}(u,y)\equiv\lim_{L\rightarrow\oo}L^{\Delta-J}\cO_{\mu_1\cdots\mu_J}(x+Ly) \ .
\end{equation}
However, instead of working with an object with indices, we can contract with another null vector $\bar{y}=(1,-\hat{\by})$ to define
\begin{equation}
    \cO(u,y)\equiv\lim_{L\rightarrow\oo}L^{\Delta-J}\bar{y}^{\mu_1}\cdots \bar{y}^{\mu_J}\cO_{\mu_1\cdots\mu_J}(x+Ly) \ .
\end{equation}
Here $\Delta$ is the mass dimension of $\cO$, $J$ is the spin of $\cO$, and $\tau\equiv\Delta-J$ is known as the twist of $\cO$. Contraction of the indices is useful for many reasons. Firstly, any traceless symmetric tensor can be written as a polynomial of $\bar{y}$ using this contraction without loss of information, by which we mean that the tensor can be fully reconstructed from the polynomial \cite{Dobrev:1977qv}. Secondly, once we integrate over future null infinity, we can analytically continue in $J$ to obtain new operators which cannot be written as the integral of a local operator. This is how the spectrum of detectors in the theory can be obtained \cite{Caron_Huot_2023}. In general, light-ray operators in conformal field theories are constructed via this notion of analytically continuing in spin \cite{Kravchuk_2018}. 

We will now integrate over $u=-x\cdot y$ . We do this in order to obtain an operator which is only a function of the direction on the celestial sphere and which models a detector which measures a physical property asymptotically far away from a scattering event.
As an example, consider choosing $\cO$ to be the stress tensor $T$. Then we find
\begin{equation}
    \int^{\oo}_{-\oo}du \ \bar{y}^{\mu}\bar{y}^{\nu}T_{\mu\nu}(u,y) \propto \cE(\hat{\by}) \ ,
\end{equation}
where $\cE(\hat{\by})$ is the well known ANEC operator \cite{Sveshnikov_1996, Tkachov_1997, Korchemsky_1999}, which measures total energy flux in a direction $\hat{\by}$ on the celestial sphere.

As we mentioned, we should try to consider detectors $\cD(\hat{\by})$ as a distribution valued operators. Let us clarify this statement somewhat now. For the purposes of this paper, which deals with scattering calculations, we will simply define this as follows: if we insert $\cD(\hat{\by})$ between any in and out scattering states in the asymptotic Fock spaces, we should obtain a well-defined distribution of functions of external momenta $p_i$:
\begin{equation}
\label{detectoropvaldist}
    \bigg|\int \frac{d^{d-1}p_1}{(2\pi)^{d-1}}\frac{1}{2\w_{p_1}}\cdots\int\frac{d^{d-1}p_N}{(2\pi)^{d-1}}\frac{1}{2\w_{p_N}}\bra{\mathrm{out}}\cD(\hat{\by})\ket{\mathrm{in}} f(p_1)\cdots f(p_N)\bigg|<\oo \ .
\end{equation}
Of course, whilst we can remove UV divergences through perturbative renormalization of field and coupling constants, the above can still fail due to IR divergences in matrix elements of the S-matrix. Nevertheless, this viewpoint is useful. Note that in general there may be additional IR divergences which require renormalization of the detector itself, which can be isolated by considering in-in event shapes with timelike external momenta, as in \cite{Caron_Huot_2023}. 

Any detector is formally defined to act on functions of external momenta via (\ref{detectoropvaldist}), where limits involved in the definition (like the $L\rightarrow\oo$ limit for our detectors) are strictly taken after integration. Nevertheless, due to various integration theorems, we may often take the limits first using a stationary phase approximation. However, it is worth noting that this should be done with caution and can lead to incorrect results if IR divergences exist.

In our theory which only has a real scalar field, the ANEC operator can also be written as \cite{Caron_Huot_2023} 
\begin{equation}
    \cE(\hat{\by})\propto\int du_1du_2|u_1-u_2|^{2(\Delta_{\phi}-1)+J_L}:\phi(u_1,y)\phi(u_2,y): \ , \hspace{0.5cm} J_L = 1-\Delta_{T} \ .
\end{equation}
Here to obtain $\cE(\hat{\by})$ we chose $J_L=1-\Delta_T$ where $\Delta_T$ is the mass dimension of the stress tensor. However, we may now analytically continue in $J_L$ to obtain the spectrum of twist-2 detectors in the theory. After renormalization at the lowest required order in perturbation theory these operators form families which are functions of a continuous real $J$. For recent work on these families in various models see \cite{Caron_Huot_2023, Henriksson:2023cnh, Li:2025knf}.

Here we have presented a working definition of some detectors and had a glimpse at how their spectrum can be obtained in a theory of a real scalar field. However, in the rest of this work we will simply define detectors as we need them, and the following sections should be self-contained.

\section{QED}
We will first discuss massive, scalar QED. Whilst this section may be seen as a warm-up before investigating the analogous story in quantum gravity, it is nevertheless physically interesting in its own right. Note that working with a scalar matter field as opposed to a fermionic matter field does not affect the physics we discuss. We use scalar fields for simplicity since all key concepts are equally well displayed in this case. Note also that we restrict to massive QED as opposed to massless QED since this is the physically relevant case (at least as far as the electromagnetic memory effect is concerned), and since it simplifies things by eliminating collinear IR singularities. 
\subsection{Classical asymptotic charges}
We begin with the case of classical, scalar, massive QED with the action
\begin{equation}
    \cS = \int d^4x\bigg[-\frac{1}{4}F^{\mu\nu}F_{\nu\mu}-m^2\phi^*\phi+(D_{\mu}\phi)^*(D^{\mu}\phi)\bigg] \ ,
\end{equation}
where $F^{\mu\nu}=\partial_{\mu}A_{\nu}-\partial_{\nu}A_{\mu}$ and  $D_{\mu}\phi = \partial_{\mu}\phi-ieA_{\mu}\phi$. One can find the equations of motion by the principle of least action, and show that these equations of motion yield a conserved current $\partial_{\mu}j^{\mu}=0$, where
\begin{equation}
    j_{\mu}=i(\phi^*\partial_{\mu}\phi-\phi\partial_{\mu}\phi^*) \ .
\end{equation}
From this we can obtain a conserved charge $Q=\int d^3x j^0$. By this we mean that everywhere along the equations of motion, $\frac{d}{dt}\int d^3x j^0=0$. Perhaps less familiar than conserved charges such as this is the weaker notion of `asymptotically conserved charges'. These can be thought of as quantities which are only conserved in the limits $t\rightarrow \pm \oo$. Thus an asymptotically conserved charge $Q$ may change with time along the equations of motion, but given a set of initial conditions the equations of motion will yield the schematic relation
\begin{equation}
     Q^{+}= Q^{-} \ ,
\end{equation}
where $Q^{+}$ is the value of $Q$ in the far future and $Q^-$ is the value of $Q$ in the far past. Since $Q$ will be written in terms of both massless and massive fields, it will have terms at both null and timelike infinities. Thus, to work with $Q^{\pm}$ we will need coordinates appropriate for working `near' these regions (where near means to some order in an $\frac{1}{r}$ expansion). Near future null infinity ($\cI^+$) we will work in retarded Bondi coordinates $(u,r,z,\bar{z})$ where $u$ is retarded time, $r$ is the spatial radius and $z$ is a complex stereographic coordinate giving the direction on the celestial sphere. Specifically, these are defined by
\begin{equation}
\label{retcoord}
    r=|\vec{x}|, \quad u =t-r, \quad x^1+ix^2 = \frac{2rz}{1+|z|^2}, \quad x^3 = r\frac{1-|z|^2}{1+|z|^2} \ .
\end{equation}
Conversely, near past null infinity ($\cI^-$) we work in advanced Bondi coordinates $(v,r,\z,\bar{\z})$ where 
\begin{equation}
\label{advcoord}
    r=|\vec{x}|, \quad v =t+r, \quad x^1+ix^2 = -\frac{2r\z}{1+|\z|^2}, \quad x^3 = -r\frac{1-|\z|^2}{1+|\z|^2} \ .
\end{equation}
Note the extra minus signs here: near $\cI^-$ we use a coordinate $\z$ which is antipodally related to $z$. The reason for this choice of convention will become clear later (it essentially just simplifies expressions). Next, near future timelike infinity $i^+$ we will use rescaled radial coordinates $(\tau,\rho,z,\bar{z})$ which can be obtained from our retarded Bondi coordinates using
\begin{equation}
    \tau = \sqrt{u^2+2ur}, \quad \rho = \frac{r}{\sqrt{u^2+2ur}} \ ,
\end{equation}
and where $z$ is defined in the same way as before. Lastly, we obtain radial rescaled coordinates $(\tau,\rho,\z,\bar{\z})$ near past timelike infinity $i^-$
from our advanced Bondi coordinates using
\begin{equation}
    \tau = \sqrt{v^2-2vr}, \quad \rho = \frac{r}{\sqrt{v^2-2vr}} \ .
\end{equation}
Again, $\z$ is defined in the same way as before. Note that for any of the four coordinate systems, our stereographic coordinate convention is such that the integration measure on the sphere is given by $d^2z\g_{z\bar{z}}$ (or $d^2\z\g_{\z\bar{\z}}$) where 
\begin{equation}
    \g_{z\bar{z}}=\frac{2}{(1+|z|^2)^2} \ .
\end{equation}
Of course, we are discussing asymptotic charges since the theory of (classical) QED contains them. With these coordinates in hand, we are ready to write down these asymptotic charges. In fact, there are infinitely many of them; in \cite{Campiglia:2015qka} it was shown that (for massive QED) one can define classical asymptotically conserved charges $Q^+_{\varepsilon}$ and $Q^-_{\varepsilon}$ using 
\begin{equation}
\label{qedclassicalcharges}
    \begin{aligned}
        &Q^{+}_{\varepsilon}=-\int_{\cI^{+}}du d^2z (\partial_z \varepsilon(z,\bar{z}) F_{u\bar{z}}+\partial_{\bar{z}}\varepsilon(z,\bar{z}) F_{uz})+e\int_{i^+} d\rho d^2z\frac{\rho^2 \gamma_{z\bar{z}}}{\sqrt{1+\rho^2}}\varepsilon_{H}(\rho,z,\bar{z})j_{\tau} \ ,\\
        &Q^{-}_{\varepsilon}=-\int_{\cI^{-}}dv d^2\z (\partial_\z \varepsilon(\z,\bar{\z}) F_{v\bar{\z}}+\partial_{\bar{\z}}\varepsilon(\z,\bar{\z}) F_{v\z})+e\int_{i^-} d\rho  d^2\z\frac{\rho^2 \gamma_{\z\bar{\z}}}{\sqrt{1+\rho^2}}\varepsilon_{H}(\rho,\z,\bar{\z})j_{\tau} \ .
    \end{aligned}
\end{equation}
Let's unpack these expressions. The terms $F_{u\bar{z}}$ and $F_{uz}$ above denote the leading terms in the large $r$ expansion at constant $u$ of the respective fields, and similarly for $F_{v\bar{z}}$ and $F_{vz}$. Likewise, the term $j_{\tau}$ is the leading order term of the  $\tau$ component of the electric current in the large $\tau$ expansion at constant $\rho$. Thus these terms are functions of the remaining coordinates $(u / v /  \rho,z/\z,\bar{z}/\bar{\z})$. These remaining coordinates parametrize what point on the boundary of Minkowski space a term approaches. The massless fields are integrated over future/past null infinity, and the massive terms are integrated over future/past timelike infinity, both against weighting functions $\varepsilon(z/\z,\bar{z}/\bar{\z})$ and $\varepsilon_{H}(\rho,z/\z,\bar{z}/\bar{\z})$ respectively. The function $\varepsilon$ is a function   $\varepsilon:S^2\rightarrow \R$ (yielding infinite possible charges). 
\footnote{We will be interested in the choice $\varepsilon(z,\bar{z}) = \gamma^{-1}_{z\bar{z}}\delta^2(z-w)$ so $\varepsilon$ is best described as a distribution on $S^2$. With this choice we don't need to worry about convergence issues at large $|z|$.} Using these, the function $\varepsilon_{H}$ is then defined by
\begin{equation}
    \begin{aligned}
        &\varepsilon_{H}(\rho,z,\bar{z})=\int \frac{d^2w}{4\pi}\frac{\varepsilon(w,\bar{w})\g_{w\bar{w}}}{(-\sqrt{1+\rho^2}+\rho \hat{\bx}_{z}\cdot \hat{\bx}_{w})^2} \ .
    \end{aligned}
\end{equation}
Here $\hat{\bx}_{z}$ is a unit 3-vector given by our stereographic convention, so
\begin{equation}
    \hat{\bx}_{z}=(\frac{z+\bar{z}}{1+|z|^2},i\frac{-z+\bar{z}}{1+|z|^2},\frac{1-|z|^2}{1+|z|^2}) \ .
\end{equation}
Then our statement that $Q_{\varepsilon}$ is asymptotically conserved translates to the following statement: given a set of initial conditions of our classical fields near $\cI^-$ and $i^-$ which specify a value of $Q^-_{\varepsilon}$, evolving along the equations of motion gives\footnote{A detailed derivation of this fact using the Lienard-Wiechert solution is given in \cite{strominger2018lecturesinfraredstructuregravity, Campiglia:2015qka}. We will not reproduce it here.}
\begin{equation}
    Q^{+}_{\varepsilon}=Q^{-}_{\varepsilon} \ .
\end{equation}
Note that for constant $\varepsilon$ we find that $Q^{\pm}_{\varepsilon}$ is proportional to the total outgoing/incoming electric charge, which we certainly expect to be conserved. For non-constant $\varepsilon$ we only expect conservation on the boundary of Minkowski space. Next we may ask if there is a similar statement we can make in the full quantum theory. Indeed there is, and we will discuss this next.
\subsection{Quantum asymptotic charges as detectors}
Moving to the quantum theory, we employ the field expansions
\begin{equation}
\begin{aligned}\label{aexp}
    &\phi(x) = \int_{p^0>0}\frac{d^4p}{(2\pi)^{3}}\delta(p^2+m^2)(e^{ipx}a(p)+e^{-ipx}b^{\dagger}(p)) \ ,\\
    &A_{\mu}(x) = \int_{p_0>0}\frac{d^dp}{(2\pi)^{d-1}}\delta(p^2)\sum^3_{r=0}(\a_r(p)\e^{r*}_{\mu}(p)e^{ipx}+\a_{r}^{\dagger}(p)\e^{r}_{\mu}(p)e^{-ipx}) \ ,
\end{aligned}
\end{equation}
where
\begin{equation}\label{ccr}
    \begin{aligned}
        &[a(p),a^{\dagger}(q)] = 2\w_p (2\pi)^{d-1}\delta^{d-1}(\bf{p}-\bf{q}) \ ,\\
        &[b(p),b^{\dagger}(q)] = 2\w_p (2\pi)^{d-1}\delta^{d-1}(\bf{p}-\bf{q}) \ ,\\
        &[\a_r(p),\a_s^{\dagger}(q)] = 2\delta_{rs}\w_p  (2\pi)^{d-1}\delta^{d-1}(\bf{p}-\bf{q}) \ .
    \end{aligned}
\end{equation}
Our first goal is to write down the charges $Q^{\pm}_{\varepsilon}$ in terms of quantum operators living at $\cI^{\pm}$ and $\cI^{\pm}_{\pm}$. An obvious guess is to try do this using detectors, which are operators which describe physical quantities measured on the boundary of Minkowski space (or at late times, such as in particle colliders). Using detectors is also a tactical choice, since they have been an active subject of research in recent years. For an overview of detectors and an introduction to the relevant literature see Section \ref{sec2}. For now, thinking of detectors as `operators describing a measurement long after a scattering process' is all we will need; in this and the next section we will give explicit definitions of detectors as we need them.  Let's start by looking at how we can put fields at timelike infinity. As an example, we could put a massless scalar field $\psi$ at $\cI^+$ by defining
\begin{equation}
    \psi(u,z)=\lim_{L\rightarrow\oo} L^{\Delta_{\psi}}\psi(x+Ly) \ ,
\end{equation}
where $u=-y\cdot x$, $\Delta_{\psi}=\frac{d-2}{2}$, and $y$ is a normalised future-pointing null vector $y^{\mu}=(1,\hat{\by})$. To ensure finiteness in the limit $L\rightarrow\oo$ for $A_{z}$, we would use
\begin{equation}
    A_z(u,y)=\lim_{L\rightarrow\oo} A_z(x+Ly) \ ,
\end{equation}
where now we have no factor of $L$ since the derivative in $z$ provides one. We can also integrate these along $\cI^+$ by integrating over $u$, and in general we may choose any weighting function of $u$ to integrate against. What is the correct weighting function for our charges? Noting that the expression $\int du F$ in our classical charge can be written as the $\w\rightarrow 0$ limit of $\int du e^{-i\w u}F$ we define, for a general local operator $X$, a so-called `$\w$-deformed' light-transform $L_{\w}[X](\oo,y)$ as
\begin{equation}
    L_{\w}[X](\oo,y) = \frac{1}{2}\int du (e^{i\w u}+e^{-i\w u})X(u,y) \ .
\end{equation}
These detectors and their properties were investigated in \cite{Korchemsky_2022}. Once the mechanism for conservation is understood, it will become clear that the Fourier transform is required so that this operator creates (soft) photons when acting on asymptotic states. Practically, we may accept this definition for now, justification will be given in the form of proof of asymptotic conservation of the charges we construct. Similar definitions follow for the case of $\cI^-$: we define
\begin{equation}
    A_{\z}(v,y)=\lim_{L\rightarrow\oo} A_{\z}(x+Ly) \ ,
\end{equation}
where $v=y\cdot x$ and $y$ is now past-pointing, so $y^{\mu}=(-1,\hat{\by})$, and 
\begin{equation}
    L_{\w}[X](-\oo,y) = \frac{1}{2}\int dv (e^{i\w v}+e^{-i\w v})X(v,y) \ .
\end{equation}
Next we want to describe how we can put our massive current $j_{\tau}$ at timelike infinity. We will use the definition
\begin{equation}
    j_{\tau}(\rho,y)=\lim_{\tau\rightarrow\oo}\tau^3 j_{\tau}(x) \ .
\end{equation}
Here $x$ is to be written using our rescaled radial coordinates described previously, and the limit $\tau$ is to be taken holding $\rho$ and $z$ constant. Using these definitions, we may now write our charges $Q^{\pm}_{\varepsilon}$ as 
\begin{equation}
\begin{aligned}
    &Q^{+}_{\varepsilon}=-\lim_{\w_0\rightarrow0}\int d^2z (\partial_z\varepsilon L_{\omega_0}[F_{u\Bar{z}}](\infty,y)+(z\leftrightarrow \bar{z}) )+e\int_{i^+} d\rho d^2z\frac{\rho^2 \gamma_{z\bar{z}}}{\sqrt{1+\rho^2}}\varepsilon_Hj_{\tau}(\rho,y) \ ,\\
    &Q^{-}_{\varepsilon}=-\lim_{\w_0\rightarrow0}\int d^2\z (\partial_\z\varepsilon L_{\omega_0}[F_{v\Bar{\z}}](-\infty,y)+(\z\leftrightarrow \bar{\z}) )+e\int_{i^-} d\rho d^2\z\frac{\rho^2 \gamma_{\z\bar{\z}}}{\sqrt{1+\rho^2}}\varepsilon_Hj_{\tau}(\rho,y) \ ,\\
\end{aligned}
\end{equation}
where we reiterate that $y$ is future-pointing if used in $Q^+_{\varepsilon}$ and past-pointing if used in $Q^-_{\varepsilon}$. The integration variable $z$ above is both the argument of $\varepsilon(z,\bar{z})$ and the stereographic coordinate of the null vector $y$; recall that the choice of $y$ determines the location on the boundary of Minkowski space that a detector is placed at. Note the limit $\w_0\rightarrow 0$ and the hidden limits $L\rightarrow \oo$ and $\tau\rightarrow \oo$. Technically, $Q^{\pm}_{\varepsilon}$ should be viewed as an operator valued distribution, and these limits are to be taken only after sandwiching $Q^{\pm}_{\varepsilon}$ between physical asymptotic scattering states and integrating against test functions of momenta, in the specific order $L\rightarrow\oo$ and $\rho\rightarrow\oo$ first and then $\w_0\rightarrow 0$. Practically, using IR finiteness of $Q^{\pm}_{\varepsilon}$ when sandwiched between dressed states and standard integration theorems we may use stationary phase approximations and take these limits before, so long as we are careful.\footnote{The order $L\rightarrow\oo$ and $\rho\rightarrow\oo$ first and then $\w_0\rightarrow 0$ should still be observed, and we should be careful not to drop soft factors which may cancel other divergences in correlation functions.} For a discussion of operator valued distributions and the formal definition of these operators see Section \ref{sec2}. As an example, the stationary phase approximation applied to $j_{\tau}(\rho,y)$ gives \cite{Campiglia:2015qka}\footnote{We have not tried to define what it means to put the actual field $\phi(x)$ at future timelike infinity. Indeed, the same definition applied to $\phi(x)$ gives a term $\sim  e^{i\tau m}$, which cannot converge when integrated against test functions of momenta in the limit $\tau\rightarrow\oo$, even if we use dressed states. Fortunately, $j_{\tau}(\rho,y)$ is all we will need.} 
\begin{equation}
    \lim_{\tau\rightarrow\oo}\tau^3j_{\tau}(x)=\frac{m^2}{2(2\pi)^3}(a_{i^+}^{\dagger}(m\rho\hat{\bx})a_{i^+}(m\rho\hat{\bx})-b_{i^+}^{\dagger}(m\rho\hat{\bx})b_{i^+}(m\rho\hat{\bx})) \ .
\end{equation}
Here and in the rest of the paper we have been careful to specify using a subscript $i^+$ that these creation and annihilation operators act on the Fock space at some constant time slice in the far future. Now we will move onto showing that $Q$ is indeed asymptotically conserved. To do this we will need to able to compute the action of the charges on asymptotic scattering states. Thus we compute the following commutators:
\begin{equation}
    \begin{aligned}
    \label{QEDCommutators1}
        &[Q^+_{\varepsilon}, a_{i^+}(p)]=-e\varepsilon_H(|\bp|/m,z_p,\bar{z}_p)a_{i^+}(p) \ ,\\
        &[Q^+_{\varepsilon}, a^{\dagger}_{i^+}(p)]=+e\varepsilon_H(|\bp|/m,z_p,\bar{z}_p)a^{\dagger}_{i^+}(p) \ ,\\
        &[Q^+_{\varepsilon}, b_{i^+}(p)]=+e\varepsilon_H(|\bp|/m,z_p,\bar{z}_p)b_{i^+}(p) \ ,\\
        &[Q^+_{\varepsilon}, b^{\dagger}_{i^+}(p)]=-e\varepsilon_H(|\bp|/m,z_p,\bar{z}_p)b^{\dagger}_{i^+}(p) \ .\\
    \end{aligned}
\end{equation}
Similarly, at $i^-$ we find
\begin{equation}
    \begin{aligned}
        &[Q^-_{\varepsilon}, a_{i^-}(p)]=-e\varepsilon_H(|\bp|/m,z_p,\bar{z}_p)a_{i^-}(p) \ ,\\
        &[Q^-_{\varepsilon}, a^{\dagger}_{i^-}(p)]=+e\varepsilon_H(|\bp|/m,z_p,\bar{z}_p)a^{\dagger}_{i^-}(p) \ ,\\
        &[Q^-_{\varepsilon}, b_{i^-}(p)]=+e\varepsilon_H(|\bp|/m,z_p,\bar{z}_p)b_{i^-}(p) \ ,\\
        &[Q^-_{\varepsilon}, b^{\dagger}_{i^-}(p)]=-e\varepsilon_H(|\bp|/m,z_p,\bar{z}_p)b^{\dagger}_{i^-}(p) \ .\\
    \end{aligned}
\end{equation}
Note that here we have denoted
\begin{equation}
    \bp = |\bp|(\frac{z_p+\bar{z}_p}{1+|z_p|^2},i\frac{-z_p+\bar{z}_p}{1+|z_p|^2},\frac{1-|z_p|^2}{1+|z_p|^2}) \ .
\end{equation}
This is consistent with the rest of this paper; the variable $z$ implies this stereographic convention whether used for detectors near $\cI^+$ or near $\cI^-$. Had we instead used the variable $\z$, the implicit assumption would've been that 
\begin{equation}
    \bp = |\bp|(-\frac{\z_p+\bar{\z}_p}{1+|\z_p|^2},-i\frac{-\z_p+\bar{\z}_p}{1+|\z_p|^2},-\frac{1-|\z_p|^2}{1+|\z_p|^2}) \ .
\end{equation}
We see from the commutation relations that when sandwiched between scattering states the effect of $Q^+_{\varepsilon}$ is to count outgoing electric charge, weighted by the function $\varepsilon_H(|\bp|/m,z,\bar{z})$ where $z$ gives the angle of the outgoing particle, and similarly for $Q^-_{\varepsilon}$ and incoming charge. Next let's compute commutators with the photon field. These are not actually needed to prove conservation but we include them for completeness. We find\footnote{Note of course that there is an implicit limit $\w_0\rightarrow 0^+$ to be taken at the end of any scattering calculation.}
\begin{equation}
\label{QcommutatorA}
    \begin{aligned}
        &[Q^+_{\varepsilon}, A_z(u,y)]=-\frac{1}{2}\partial_{z}\varepsilon(z_y,\bar{z}_y) \ ,\\
        &[Q^+_{\varepsilon}, A_{\bar{z}}(u,y)]=-\frac{1}{2}\partial_{\bar{z}}\varepsilon(z_y,\bar{z}_y) \ ,\\
        &[Q^+_{\varepsilon},\a_{1,\cI^+}(p)] = -\sqrt{2}\pi^2(1+|z_p|^2)\delta(\w_p-\w_0)\partial_{\bar{z}}\varepsilon(z_p,\bar{z}_p) \ ,\\
        &[Q^+_{\varepsilon},\a_{2,\cI^+}(p)] = -\sqrt{2}\pi^2(1+|z_p|^2)\delta(\w_p-\w_0)\partial_{z}\varepsilon(z_p,\bar{z}_p) \ ,\\
        &[Q^+_{\varepsilon},\a_{1,\cI^+}^{\dagger}(p)] = \sqrt{2}\pi^2(1+|z_p|^2)\delta(\w_p-\w_0)\partial_{z}\varepsilon(z_p,\bar{z}_p) \ ,\\
        &[Q^+_{\varepsilon},\a_{2,\cI^+}^{\dagger}(p)] = \sqrt{2}\pi^2(1+|z_p|^2)\delta(\w_p-\w_0)\partial_{\bar{z}}\varepsilon(z_p,\bar{z}_p) \ ,\\
    \end{aligned}
\end{equation}
and similarly for $Q^-_{\varepsilon}$. For a discussion of the factor of $\frac{1}{2}$ in front of the first two commutators, see Appendix \ref{AppendixA}. All that is left is to discern the effect of $Q^{\pm}_{\varepsilon}$ on the left and right vacuums. To investigate this, we can use a stationary phase approximation of $Q^{\pm}_{\varepsilon}$ itself (recalling again that we should be careful when using this). Let us write
\begin{equation}
    Q^{+}_{\varepsilon} = Q^{+}_{\varepsilon, \ \mathrm{soft}} + Q^{+}_{\varepsilon, \ \mathrm{hard}},
\end{equation}
where 
\begin{equation}
    \begin{aligned}
        &Q^{+}_{\varepsilon, \ \mathrm{soft}} = -\lim_{\w_0\rightarrow0}\int d^2z (\partial_z\varepsilon(z,\bar{z}) L_{\omega_0}[F_{u\Bar{z}}](\infty,y)+(z\leftrightarrow \bar{z}) ) \ ,\\
        &Q^{+}_{\varepsilon, \ \mathrm{hard}} = e\int_{i^+} d\rho d^2z\frac{\rho^2 \gamma_{z\bar{z}}}{\sqrt{1+\rho^2}}\varepsilon_H(\rho,z,\bar{z})j_{\tau}(\rho,y) \ ,
    \end{aligned}
\end{equation}
and similarly for $Q^{-}_{\varepsilon}$. Let's start by looking at the stationary phase approximation of simpler detectors. Consider $A_{z/\bar{z}}(u,y)=\lim_{L\rightarrow \oo}A_{z/\bar{z}}(x+Ly)$ where $y$ is future pointing, $y^{\mu}=(1,\hat{\by})$. The stationary phase approximation gives
\begin{equation}
    A_{z/\bar{z}}(u,y) = \frac{i}{2(2\pi)^2}\int^{\oo}_0d\w(e^{i\w u}\sum_{r=1,2}\e^{r}_{z/\bar{z}}(y)\a^{\dagger}_{r,\cI^+}(\w y)-e^{-i\w u}\sum_{r=1,2}\e^{r*}_{z/\bar{z}}(y)\a_{r,\cI^+}(\w y)) \ ,
\end{equation}
where $y=(0,1,z,\bar{z})$ in retarded Bondi coordinates $(u,r,z,\bar{z})$ and we have chosen our polarization vectors according to
\begin{equation}
    \begin{aligned}
        \e^1(y) = \frac{1}{\sqrt{2}}(\bar{z},1,-i,-\bar{z}), \quad \e^2(y) = \frac{1}{\sqrt{2}}(z,1,i,-z) \ .
    \end{aligned}
\end{equation}
This gives
\begin{equation}
\begin{aligned}
    &\e^{1}_{z} = \frac{1}{\sqrt{2}}(\frac{(1-\bar{z}^2)}{(1+z\bar{z})^2}-\frac{(1+\bar{z}^2)}{(1+z\bar{z})^2}+\frac{2\bar{z}^2}{(1+z\bar{z})^2})= 0 \ ,\\
    &
\e^1_{\bar{z}} = \frac{1}{\sqrt{2}}(\frac{(1-z^2)}{(1+z\bar{z})^2}+\frac{(1+z^2)}{(1+z\bar{z})^2}+\frac{2z\bar{z}}{(1+z\bar{z})^2})=\frac{\sqrt{2}}{1+z\bar{z}} \ .
\end{aligned}
\end{equation}
Similarly, $\e^2_z =\frac{\sqrt{2}}{1+z\bar{z}}$ and $\e^2_{\bar{z}}=0$. Substituting in we find
\begin{equation}
    \begin{aligned}
        A_{z}(u,y) &= \frac{i}{\sqrt{2}(2\pi)^2(1+|z|^2)}\int^{\oo}_0d\w(e^{i\w u}\a_{2,\cI^+}^{\dagger}(\w y)-e^{-i\w u}\a_{1,\cI^+}(\w y)) \ ,\\
        A_{\bar{z}}(u,y) &= \frac{i}{\sqrt{2}(2\pi)^2(1+|z|^2)}\int^{\oo}_0d\w(e^{i\w u}\a_{1,\cI^+}^{\dagger}(\w y)-e^{-i\w u}\a_{2,\cI^+}(\w y)) \ .
    \end{aligned}
\end{equation}
The effect of taking $L_{\w_0}[A_z](\oo,y)$ simply undoes the Fourier transform. If we had to use $\partial_{u}A_z$ instead, we would get a factor of the momentum in front. Altogether we find that $Q^{+}_{\varepsilon, \ \mathrm{soft}}$ is given by
\begin{equation}
    \lim_{\w_0\rightarrow 0}\frac{\w_0}{8\pi}\int d^2z\sqrt{\g_{z\bar{z}}}\bigg(\partial_z \varepsilon(\a_{1,\cI^+}^{\dagger}(\w_0 y)+\a_{2,\cI^+}(\w_0y))+\partial_{\bar{z}} \varepsilon(\a_{2,\cI^+}^{\dagger}(\w_0 y)+\a_{1,\cI^+}(\w_0y))\bigg) \ .
\end{equation}
The effect of this on the vacuum is to create soft photons, hence the name. Via the soft theorem, these will lead to divergences in $\w_0$, which are in turn canceled by the factor of $\w_0$ in the numerator of $Q^{+}_{\varepsilon, \ \mathrm{soft}}$. Furthermore, noting that the creation and annihilation operators in $j_{\tau}(\rho,y)$ are normal ordered tells us that $Q^{+}_{\varepsilon, \ \mathrm{hard}}$ simply annihilates the left and right vacuum. Obviously, similar results hold for $Q^-_{\varepsilon}$. Thus, we now know enough to show asymptotic conservation, which we do in the next subsection.
\subsection{Asymptotic conservation for undressed external states}
Now we will show that for arbitrary scattering states $\bra{\mathrm{out}}$ and $\ket{\mathrm{in}}$ we have 
\begin{equation}
\label{qedundressed1}
    \bra{\mathrm{out}}Q^+_{\varepsilon}\ket{\mathrm{in}}-\bra{\mathrm{\mathrm{out}}}Q^-_{\varepsilon}\ket{\mathrm{in}}= 0 \ .
\end{equation}
This was first shown for massive QED in \cite{Campiglia:2015qka}. In particular, this follows from the soft theorem associated with QED.\footnote{By undressed, we mean states without a Faddeev-Kulish dressing as defined in \cite{Kulish:1970ut}. These are the normal asymptotic momentum scattering states created by creation and annihilation operators in canonical quantization.} We begin by considering the case where all external particles are scalar, but we will be able to immediately generalize to any scattering process. Thus, let the $\ket{\mathrm{in}}$ state consists of $n$ scalar particles and the $\bra{\mathrm{out}}$ state of $m$ scalar particles, with momenta and charges indexed by $s\in\{1,...,n\}$ and $s'\in\{1,...,m\}$ respectively. Recall that we have
\begin{equation}
    \begin{aligned}
        &[Q^{+}_{\varepsilon},a_{i^+}(p_{s'})]= -e\varepsilon_H(\frac{|\bp_{s'}|}{m},z_{s'},\bar{z}_{s'})a_{i^+}(p_{s'}) \ ,\\
        &[Q^{+}_{\varepsilon},b_{i^+}(p_{s'})]= +e\varepsilon_H(\frac{|\bp_{s'}|}{m},z_{s'},\bar{z}_{s'})b_{i^+}(p_{s'}) \ .
    \end{aligned}
\end{equation}
and 
\begin{equation}
    \begin{aligned}
        &[Q^{-}_{\varepsilon},a^{\dagger}_{i^-}(p_{s})]= +e\varepsilon_H(\frac{|\bp_{s'}|}{m},z_{s},\bar{z}_{s})a^{\dagger}_{i^-}(p_{s}) \ ,\\
        &[Q^{-}_{\varepsilon},b^{\dagger}_{i^-}(p_{s})]= -e\varepsilon_H(\frac{|\bp_{s'}|}{m},z_{s},\bar{z}_{s})b^{\dagger}_{i^-}(p_{s}) \ .
    \end{aligned}
\end{equation}
Substituting these into (\ref{qedundressed1}) we immediately find
\begin{equation}
\begin{aligned}
    &\bra{\mathrm{out}}Q^+_{\varepsilon}\ket{\mathrm{in}} - \bra{\mathrm{out}}Q^-_{\varepsilon}\ket{\mathrm{in}}\\
    =&\bra{\mathrm{out}}Q^+_{\varepsilon, \ \mathrm{soft}}\ket{\mathrm{in}} -\bra{\mathrm{out}} Q^-_{\varepsilon, \ \mathrm{soft}}\ket{\mathrm{in}}\\&-\bra{\mathrm{out}}\ket{\mathrm{in}}\bigg(\sum_{s'=1}^{m}q_{s'}\varepsilon_H(\frac{|\bp_{s'}|}{m}z_{s'},\bar{z}_{s'})-\sum_{s=1}^{n}q_{s}\varepsilon_H(\frac{|\bp_{s'}|}{m}z_{s},\bar{z}_{s})\bigg) \ .
\end{aligned}
\end{equation}
Here $z_{s'/s}$ is the stereographic coordinates associated with the momentum $p_{s'/s}$ and the electric charges $q_{s/s'}$ can take values $\pm e$. It remains to calculate the contribution from $Q^{\pm}_{\varepsilon, \ \mathrm{soft}}$. We may use the stationary phase approximation here to write $\bra{0}Q^+_{\varepsilon, \ \mathrm{soft}}$ as
\begin{equation}
    \begin{aligned}
        \bra{0} \frac{\w_0}{8\pi}\int d^2z\sqrt{\gamma_{z\bar{z}}} \bigg(\partial_z \varepsilon(\a_{1,\cI^+}^{\dagger}(\w_0 y)+\a_{2,\cI^+}(\w_0y))+\partial_{\bar{z}} \varepsilon(\a_{2,\cI^+}^{\dagger}(\w_0 y)+\a_{1,\cI^+}(\w_0y))\bigg) \ .
    \end{aligned}
\end{equation}
Thus, using the soft theorem, we find
\begin{equation}
    \begin{aligned}
        &\bra{\mathrm{out}}Q^+_{\varepsilon, \ \mathrm{soft}}\ket{\mathrm{in}} \\ =&\int d^2z \partial_{\bar{z}}\varepsilon(z,\bar{z})\frac{1}{8\pi }|\w_0|\sqrt{\gamma_{z\bar{z}}}\bigg(\sum^{n}_{s=1}\frac{q_sp_s\cdot\e^{1}(z,\bar{z})}{|\w_0|p_s\cdot y}-\sum^{m}_{s'=1}\frac{q_{s'}p_{s'}\cdot\e^{1}(z,\bar{z})}{|\w_0|p_{s'}\cdot y}\bigg)\braket{\mathrm{out}}{\mathrm{in}}\\
        &+ \int d^2z \partial_z\varepsilon(z,\bar{z})\frac{1}{8\pi }|\w_0|\sqrt{\gamma_{z\bar{z}}}\bigg(\sum^{n}_{s=1}\frac{q_sp_s\cdot\e^{2}(z,\bar{z})}{|\w_0|p_s\cdot y}-\sum^{m}_{s'=1}\frac{q_{s'}p_{s'}\cdot\e^{2}(z,\bar{z})}{|\w_0|p_{s'}\cdot y}\bigg)\braket{\mathrm{out}}{\mathrm{in}} \ .
    \end{aligned}
\end{equation}
It is not hard to show that 
\begin{equation}
    \partial_{\bar{z}}\bigg(\frac{\sqrt{\gamma_{z\bar{z}}}p_{s'}\cdot \e^1(z,\bar{z})}{p_{s'}\cdot y}\bigg)=\frac{\g_{z\bar{z}}}{2(-\sqrt{1+\rho^2_{s'}}+\rho_{s'}\hat{\bx}_{s'}\cdot \hat{\bx}_{z})^2} \ .
\end{equation}
Thus, integration by parts gives
\begin{equation}
    \begin{aligned}
        \bra{\mathrm{out}}Q^+_{\varepsilon, \ \mathrm{soft}}\ket{\mathrm{in}} =\frac{1}{2}\bigg(-\sum^{n}_{s=1}q_s\varepsilon_H(\rho_s,z_s,\bar{z}_s)+\sum^{m}_{s'=1}q_{s'}\varepsilon_H(\rho_{s'},z_{s'},\bar{z}_{s'})\bigg)\braket{\mathrm{out}}{\mathrm{in}} \ .
    \end{aligned}
\end{equation}
Similarly, we find
\begin{equation}
    \begin{aligned}
        \bra{\mathrm{out}}Q^-_{\varepsilon, \ \mathrm{soft}}\ket{\mathrm{in}} =\frac{1}{2}\bigg(\sum^{n}_{s=1}q_s\varepsilon_H(\rho_{s},z_s,\bar{z}_s)-\sum^{m}_{s'=1}q_{s'}\varepsilon_H(\rho_{s'},z_{s'},\bar{z}_{s'})\bigg)\braket{\mathrm{out}}{\mathrm{in}} \ .
    \end{aligned}
\end{equation}
Substituting this immediately gives the required
\begin{equation}
    \bra{\mathrm{out}}Q^+_{\varepsilon}\ket{\mathrm{in}}-\bra{\mathrm{out}}Q^-_{\varepsilon}\ket{\mathrm{in}}=0 \ .
\end{equation}
Note that we obtained asymptotic conservation from the soft theorem despite our factor of $\frac{1}{2}$ in the commutator $[Q^{+}_{\varepsilon},A_z]$, since it was not required for this calculation. Nevertheless, it appears that one needs the factor of $\frac{1}{2}$ to ensure conservation of $Q$ if one uses FK dressed states. For more on this note see Appendix \ref{AppendixA}. Next, what about if we also have external hard photons? Using, for example,
\begin{equation}
    \begin{aligned}
        &[Q^+_{\varepsilon, \ \mathrm{soft}},\a_{1,\cI^+}(p)] = -\sqrt{2}\pi^2(1+|z_p|^2)\delta(\w_p-\w_0)\partial_{\bar{z}}\varepsilon(z_p,\bar{z}_p) \ ,
    \end{aligned}
\end{equation}
we find, 
\begin{equation}
    \bra{\mathrm{out}}\a_{1,\cI^+}(p)Q^+_{\varepsilon}\ket{\mathrm{in}}\propto\sqrt{2}\pi^2(1+|z_p|^2)\delta(\w_p-\w_0)\braket{\mathrm{out}}{\mathrm{in}}  \ .
\end{equation}
Since we must eventually take the limit $\w_0\rightarrow 0^+$ this is the zero distribution on Schwartz functions of null momenta $p$ (note that here $\braket{\mathrm{out}}{\mathrm{in}}$ is independent of $p$). This finally confirms that 
\begin{equation}
    \bra{\mathrm{out}}Q^+_{\varepsilon}\ket{\mathrm{in}}-\bra{\mathrm{out}}Q^-_{\varepsilon}\ket{\mathrm{in}}=0 \ .
\end{equation}
holds for all possible (undressed) $\bra{\mathrm{out}}$ and $\ket{\mathrm{in}}$. Since we required the soft theorem to show this, we can say that this asymptotic conservation is in fact implied by the soft theorem (and vice-versa). This subsection has served to simply confirm the result obtained in \cite{Campiglia:2015qka}.
\subsection{Asymptotic conservation for dressed external states} 
In the preceding section we used the fact that scattering amplitudes in QED suffer from IR divergences when external photons become soft. These poles, described by the soft theorem, were multiplied by zeroes in the soft part of $Q^{\pm}_{\varepsilon}$. This led to finite terms which cancel those introduced by the hard part of $Q^{\pm}_{\varepsilon}$. However, building on earlier work \cite{PhysRev.140.B1110} by Chung, a paper by Faddeev and Kulish in 1970 constructed `physical' asymptotic particle states such that these scattering calculations do not suffer from these IR divergences \cite{Kulish:1970ut}. Here we will show that using these FK dressed states upholds conservation of $Q^{\pm}_{\varepsilon}$ in the sense that for any scattering process
\begin{equation}
    \begin{aligned}
        &\bra{\mathrm{out}, \  \mathrm{dressed}}Q^+_{\varepsilon}\ket{\mathrm{in}, \ \mathrm{dressed}}-\bra{\mathrm{out}, \  \mathrm{dressed}}Q^-_{\varepsilon}\ket{\mathrm{in}, \ \mathrm{dressed}}=0 \ .
    \end{aligned}
\end{equation}
This was first shown to hold in \cite{Gabai_2016}. As in \cite{Gabai_2016} we will also find that when the net electric charge is zero
\begin{equation}
    \begin{aligned}
        &\bra{\mathrm{out}, \  \mathrm{dressed}}Q^+_{\varepsilon}\ket{\mathrm{in}, \ \mathrm{dressed}}=\bra{\mathrm{out}, \  \mathrm{dressed}}Q^-_{\varepsilon}\ket{\mathrm{in}, \ \mathrm{dressed}}=0 \ .
    \end{aligned}
\end{equation} 
For general scattering, we find that conservation is trivialised in the sense that
\begin{equation}
    \bra{\mathrm{out}, \  \mathrm{dressed}}Q^+_{\varepsilon}\ket{\mathrm{in}, \ \mathrm{dressed}} = \sum_{s'}q_{s'}\frac{1}{4\pi}\int d^2z \gamma_{z\bar{z}}\varepsilon(z,\bar{z}) \ \braket{\mathrm{out}, \  \mathrm{dressed}}{\mathrm{in}, \ \mathrm{dressed}} \ , 
\end{equation}
which is independent of the momenta of the external particles, as shown in \cite{Gabai_2016}. In the next subsection we will discuss how this result relates to the memory effect. To show that it holds let us begin by discussing how to construct FK dressed states. FK states involve multiplying our asymptotic states by an operator which is designed to cancel all IR divergences from S-matrix calculations. Given any in and out asymptotic scattering states $\bra{\mathrm{out}}$ and $\ket{\mathrm{in}}$ we define \cite{Kulish:1970ut}
\begin{equation}
    \begin{aligned}
        &\ket{\mathrm{in}, \ \mathrm{dressed}} = e^{-R_{f}}\ket{\mathrm{in}} \ ,\\
        &\bra{\mathrm{out}, \ \mathrm{dressed}} = \bra{\mathrm{out}}e^{R_{f}} \ ,
    \end{aligned}
\end{equation}
where
\begin{equation}
    R_f = \int \frac{d^3 \bp}{(2\pi)^3}\frac{\rho(\bp)}{2\w_p}\int\frac{d^3\bk}{(2\pi)^3}\frac{1}{2\w_k}[f(k,p)\cdot \e^{r}\a^{\dagger}_{r}(k)-f^*(k,p)\cdot\e^{r*}(k)\a_{r}(k)] \ .
\end{equation}
Here
\begin{equation}
    \rho(\bp) = e(a^{\dagger}(p)a(p)-b^{\dagger}(p)b(p)) \ ,
\end{equation}
and\footnote{Later in the gravity case we will explicitly include the required $t$ dependence of the dressing in order to obtain the correct results. We will not need this here since photons do not self-interact. Note, however, that it is technically required to describe the subleading $1/t$ behaviour as will become clear in the gravity case.} 
\begin{equation}
    f_{\mu}(k,p) = (\frac{p_{\mu}}{k\cdot p}-\frac{c_{\mu}}{\w_k})\varphi(k,p) \ ,
\end{equation}
where $\varphi(k,p)$ is a smoothing function satisfying $\varphi(k,p)=1$ in a neighbourhood of $k=0$. We have freedom in choosing the $k$ dependent null vector $c_{\mu}(k)$. Specifically, we can write $c_{\mu}$ as
\begin{equation}
    c_{\mu}= \frac{\w_k q(k)_{\mu}}{k\cdot q(k)} \ ,
\end{equation}
where the necessary conditions outlined in \cite{Kulish:1970ut} are satisfied if $q(k)$ is any map from null vectors $k$ to a null vector $q(k)\neq k$, where $q(k)$ is homotopic to the antipodal map. As was done in \cite{Kulish:1970ut} we will use the unique rotationally invariant choice given by $q(k)=(1,-\hat{\bk})$. The point is that after commuting $e^{\pm R_f}$ with the asymptotic creation and annihilation operators each external matter particle is dressed with a cloud of infinite photons, which correspond to the actual `physical states' that we measure. Now we will show that $Q^{\pm}_{\varepsilon}$ is asymptotically conserved if one uses Faddeev-Kulish dressed external states, by which we mean
\begin{equation}
    \begin{aligned}
        &\bra{\mathrm{out}, \  \mathrm{dressed}}Q^+_{\varepsilon}\ket{\mathrm{in}, \ \mathrm{dressed}}-\bra{\mathrm{out}, \  \mathrm{dressed}}Q^-_{\varepsilon}\ket{\mathrm{in}, \ \mathrm{dressed}}=0 \ .
    \end{aligned}
\end{equation}
First, note that due to the factor $\rho(\bp)$
\begin{equation}
    \begin{aligned}
        &[a(p),R_f]=e \ a(p)\int\frac{d^3\bk}{(2\pi)^3}\frac{1}{2\w_k}[f(k,p)\cdot \e^{*r}\a^{\dagger}_{r}(k)-f^*(k,p)\cdot\e^{r}(k)\a_{r}(k)] \ .
    \end{aligned}
\end{equation}
Thus, for each external scalar particle in $\bra{\mathrm{out}, \  \mathrm{dressed}}$ with momentum $p$ and charge $q=\pm e$ there is an exponential factor $e^{iq\Phi(p)}$, where we have defined
\begin{equation}
\label{FKphidefqed}
    \Phi(p)=-i\int\frac{d^3\bk}{(2\pi)^3}\frac{1}{2\w_k}[f(k,p)\cdot \e^{*r}\a^{\dagger}_{r}(k)-f^*(k,p)\cdot\e^{r}(k)\a_{r}(k)] \ .
\end{equation}
This is the factor which `dresses' this external hard particle with a cloud of photons. Next, we calculate
\begin{equation}
\label{Fkdressingcommutatorqed}
    \begin{aligned}
        &[Q^+_{\varepsilon, \ \mathrm{soft}},\Phi(p)] =i\varepsilon_H(\rho_p,z_{p},\bar{z}_{p})+\frac{i}{4\pi}\int d^2z \gamma_{z\bar{z}}\varepsilon(z,\bar{z}) \ ,\\
        &[Q^-_{\varepsilon, \ \mathrm{soft}},\Phi(p)] =i\varepsilon_H(\rho_p,z_{p},\bar{z}_{p})+\frac{i}{4\pi}\int d^2z \gamma_{z\bar{z}}\varepsilon(z,\bar{z}) \ .
    \end{aligned}
\end{equation}
For a discussion of these commutators as they relate to the factor of $\frac{1}{2}$ in (\ref{QcommutatorA}) see Appendix \ref{AppendixA}. In particular, we use the standard commutation relations which include the factor of $\frac{1}{2}$ in (\ref{QcommutatorA}) to obtain the commutators in (\ref{Fkdressingcommutatorqed}), which in turn are required for conservation of $Q$.  Returning to our calculation, note that the first terms on the RHS of each line in (\ref{Fkdressingcommutatorqed}) exactly cancel the terms arising from commuting $Q^{\pm}_{\varepsilon}$ with the matter creation and annihilation operators, whilst the second terms are independent of the momenta of external particles. Lastly, all we need to note is that since dressed scattering states are free of IR divergences, $Q^{\pm}_{\varepsilon}$ now annihilates the vacuum when commuted past everything, since there are no divergences to cancel the soft factor.\footnote{At least $\bra{0}Q^+_{\varepsilon}$ gives zero when sandwiched with any incoming dressed ket. We may then formally take $\bra{0}Q^+_{\varepsilon}=0$ if we agree to work with FK dressed asymptotic Fock spaces and study the associated S-matrix. } Thus, given scattering states $\bra{\mathrm{out}, \  \mathrm{dressed}}$ and $\ket{\mathrm{in}, \  \mathrm{dressed}}$ which consists of any number of external hard matter particles and photons, we immediately have
\begin{equation}
    \begin{aligned}
        &\bra{\mathrm{out}, \  \mathrm{dressed}}Q^+_{\varepsilon}\ket{\mathrm{in}, \ \mathrm{dressed}}-\bra{\mathrm{out}, \  \mathrm{dressed}}Q^-_{\varepsilon}\ket{\mathrm{in}, \ \mathrm{dressed}}\\
        =&(\sum_{s'}q_{s'}-\sum_{s}q_{s})\frac{1}{4\pi}\int d^2z \gamma_{z\bar{z}}\varepsilon(z,\bar{z})\\
        =&0 \ .
    \end{aligned}
\end{equation}
In particular note that 
\begin{equation}
    \begin{aligned}
        \bra{\mathrm{out}, \  \mathrm{dressed}}Q^+_{\varepsilon}=\sum_{s'}q_{s'}\frac{1}{4\pi}\int d^2z \gamma_{z\bar{z}}\varepsilon(z,\bar{z}) \ \bra{\mathrm{out}, \  \mathrm{dressed}} \ .  
    \end{aligned}
\end{equation}
Thus FK dressings allow us to construct a basis of the Fock space which is made up of eigenvectors of $Q^{^+}_{\varepsilon}$, and the eigenvalues are independent of external momenta. In the next section we will interpret this result in terms of physical asymptotic Fock spaces and the memory effect.
\subsection{Physical asymptotic Fock spaces and the memory effect}
The previous two subsections were simply dedicated to defining our asymptotic charges and showing the trivialization of their conservation when sandwiched between FK dressed scattering states. We reiterate that these results are obtained using rigorous definitions of our detectors as limiting to well-defined operator valued distributions.\footnote{Of course, whilst there are no IR divergences, UV divergences still exist and appear in the overlaps $\braket{\mathrm{out}}{\mathrm{in}}$, but it is understood how to deal with these using standard renormalization theory.} Now we will briefly discuss the physical interpretation of these results. To do so we must consider the memory effect in QED and how it leads to the IR divergences in S-matrix elements which the FK dressing removes. For a more complete explanation of this last point, see the introduction of \cite{Prabhu:2022zcr}, which some of this discussion will be based on. 

The memory effect in QED describes the important physical phenomenon where, to leading order in $\frac{1}{r}$, the angular components of the electromagnetic potential $A$ at late retarded times differ from their values at early retarded times. Since a fully consistent and well-defined construction of the asymptotic Fock spaces in massive QED must capture all relevant physics, they must encode the memory effect. However, attempting to encode memory using the standard construction of asymptotic Fock spaces does not work. This is because in $d=4$ dimensions the memory effect and radiation both decay as $1/r$ \cite{Satishchandran_2019}. Encoding the memory effect requires adding soft photons to the in and out asymptotic spaces, but this leads to IR divergences described by the soft theorem.

The famous IR divergences in the massive QED S-matrix can thus be interpreted as occurring due to the failure of the Fock space to include the memory effect. We have seen that FK dressings remove these divergences and simultaneously trivialize the conservation of $Q^{\pm}_{\varepsilon}$ which implies a relation between our charges $Q^{\pm}_{\varepsilon}$ and the electromagnetic memory. Indeed, this is known \cite{Bieri_2013, Prabhu:2022zcr}. To find it, let $A_{X}^{(1)}(u,z,\bar{z})$ be the leading ($\frac{1}{r}$) term of the $X\in\{z,\bar{z}\}$ component of the electromagnetic potential near $\cI^+$. Then, the leading term of the electric field is given by 
\begin{equation}
    E_{X}^{(1)} = -\partial_{u}A^{(1)}_{X} \ ,
\end{equation}
and by definition we have a non-zero memory effect at $\cI^+$ if \cite{Bieri_2013}
\begin{equation}
    \Delta_X\equiv\int_{-\oo}^{\oo}du E^{(1)}_{X}=A^{(1)}_{X}|_{u=+\oo}-A^{(1)}_{X}|_{u=-\oo}\neq 0 \ .
\end{equation}
Here $\Delta_X$ is the `memory out' and depends on both the incoming state and the outgoing state. In particular Bieri and Garfinkle find \cite{Bieri_2013}
\begin{equation}
    D^X\Delta_X=\bigg(E^{(2)}_{r}(\oo)-[E^{(2)}_{r}(\oo)]_{[0]}\bigg)-\bigg(E^{(2)}_r(-\oo)-[E^{(2)}_r(-\infty)]_{[0]}\bigg) \ ,
\end{equation}
where we have dropped the null current term since we have no massless charged particles in our theory,  $E^{(2)}_{r}(u)$ is the leading term of the radial component of the electric field in a $1/r$ expansion at retarded time $u$. Note that $[E_r(\oo)]_{[0]}$ denotes the average of $E_r(\oo)$ when integrated over the sphere. Here the first two terms in brackets are the contribution from the outgoing state, and the last two are the contribution from the incoming state.

Next, if we consider the first term in (\ref{qedclassicalcharges}), which we called $Q^+_{\varepsilon, \ \mathrm{soft}}$, we see after a little algebra that this is actually related to the memory by \cite{Pasterski:2015zua}
\begin{equation}
    Q^+_{\varepsilon, \ \mathrm{soft}} \equiv \int d\Omega \  D^{X}\varepsilon \ \Delta_X \ .
\end{equation}
It is important to note that the memory due to the outgoing state is not dependent on the memory due to the ingoing state. A consistent, physical Fock space should thus only encode the memory due to the outgoing state in the outgoing Fock space. Indeed, this is exactly what we find. To see this first note that one should really be working with the operator $Q^+_{\hat{y}, \ \mathrm{soft}}$ where $\hat{y}$ labels a point on the celestial sphere. This is equivalent to choosing 
\begin{equation}
    \varepsilon(z,\bar{z}) = \frac{1}{\gamma_{z\bar{z}}}\delta^2(z-z_y) \ .
\end{equation}
We then have 
\begin{equation}
    Q^+_{\hat{y}, \ \mathrm{soft}}=-D^X\Delta_{X}(z,\bar{z}) \ .
\end{equation}
So $-Q^+_{\hat{y}, \ \mathrm{soft}}$ measures the memory due to the outgoing state at a point at a point on the celestial sphere labeled by $\hat{y}$. Indeed one finds
\begin{equation}
    \bra{\mathrm{out}, \ \mathrm{dressed}}Q^+_{\hat{y}, \ \mathrm{soft}}= \bigg(\frac{1}{4\pi}\sum_{s'}\frac{q_{s'}p^2_{s'}}{(p_{s'}\cdot \hat{y})^2}+\frac{1}{4\pi}\sum_{s'}q_{s'}\bigg)\bra{\mathrm{out}, \ \mathrm{dressed}} \ .
\end{equation}
This is exactly the prediction by Bieri and Garfinkle. Indeed it is not hard to show that given our choice of units we have
\begin{equation}
    E^{(2)}_{r}(\oo)=-\frac{1}{4\pi}\sum_{s'}\frac{q_{s'}p^2_{s'}}{(p_{s'}\cdot \hat{y})^2} \ , \quad[E^{(2)}_{r}(\oo)]_{[0]}=\frac{1}{4\pi}\sum_{s'}q_{s'} \ .
\end{equation}
Thus we see that there is an additional constant contribution equal to the total charge of the process. Of course, due to charge conservation one has 
\begin{equation}
    [E^{(2)}_{r}(\oo)]_{[0]}=[E^{(2)}_{r}(-\oo)]_{[0]} \ ,
\end{equation}
and so this term does cancel when computing the full memory. However, we argue that it should be included in the eigenvalue since it arises from the $c_{\mu}$ term in the dressing which is required to gauge fix the Fock space. Thus we do expect physical state vectors to contain this contribution to the eigenvalue.
 Note that this term then cannot be `large gauge transformed away'.
 
 Note that the dressed states being eigenstates of $Q^{+}_{\varepsilon \ \mathrm{soft}}$ implies that the FK dressings encode the outgoing memory effect into the outgoing states, and similarly for the incoming memory and states. Thus removing the IR divergences is equivalent to encoding the memory correctly. This may suggest that the dressed Fock spaces may be considered to be more physical than the undressed spaces. One may define this physical Fock space as eigenvalues of the operator $Q^+_{
\hat{y}}$, where any state satisfies 
\begin{equation}
    \bra{\mathrm{out}, \ \mathrm{dressed}}Q^+_{\hat{y}}= \frac{1}{4\pi}\sum_{s'}q_{s'}\bra{\mathrm{out}, \ \mathrm{dressed}} \ .
\end{equation}
Due to the discussion in \cite{Prabhu:2022zcr} it may be worth noting that this eigenvalue is Lorentz invariant and independent of $\hat{y}$. 

\section{Perturbative quantum gravity}
\subsection{Classical asymptotic charges}
\subsubsection{Asymptotically flat spacetimes}
Now we wish to tell the same story in perturbative quantum gravity. As before we begin here by describing the asymptotically conserved charges in the classical theory. Our discussion of the classical theory is heavily based on the more complete discussions in \cite{He_2015, strominger2018lecturesinfraredstructuregravity}; our only goal here is to state the existence of asymptotically conserved charges which we will rigorously define and work with in the quantum theory in the next subsection.  We consider the Einstein-Hilbert action coupled to a massive real scalar:
\begin{equation}
    S = \int d^4x\sqrt{-g}\bigg(\frac{1}{16\pi G}R+\frac{1}{2}g^{\mu\nu}\partial_{\mu}\phi\partial_{\nu}\phi-\frac{1}{2}m^2\phi^2\bigg) \ .
\end{equation}
Expanding using $g_{\mu\nu}=\eta_{\mu\nu}+\sqrt{32\pi G}h_{\mu\nu}$
gives a Lagrangian consisting of a spin-2 massless field $h_{\mu\nu}$ (the graviton) coupled to our real scalar $\phi$. Varying this action gives the Einstein-Hilbert equations, which may be used to find solutions on $\cI^+$ given a set of initial data on $\cI^-$. We will be interested in the case of asymptotically flat metrics, so our initial and final metric is flat up to terms which decay for large radius $r$. Specifically, we choose our coordinate gauge and falloff conditions according to Bondi, van der Burg, Metzner, and Sachs \cite{1962RSPSA.269...21B, 1962RSPSA.270..103S}. Using the retarded Bondi coordinates defined in (\ref{retcoord}), these conditions constrain the metric near $\cI^+$ to be
\begin{equation}
\begin{aligned}
ds^2 =& -du^2 -2dudr +2r^2\gamma_{z\Bar{z}}dzd\Bar{z}+\frac{2m_B}{r}du^2+rC_{zz}dz^2+rC_{\Bar{z}\Bar{z}}d\Bar{z}^2\\&+D^zC_{zz}dudz+D^{\Bar{z}}C_{\Bar{z}\Bar{z}}dud\Bar{z}
+... \  ,
\end{aligned}
\end{equation}
where $D_z$ is the covariant derivative with respect to the metric on the unit sphere $\gamma_{z\Bar{z}}$. Note that $C_{zz}$ and $m_B$  do not depend on $r$. To obtain this metric we enforced the following falloffs in $r$:
\begin{equation}
\begin{aligned}
&g_{uu} = -1 + \cO(\frac{1}{r}), \hspace{0,5 cm} g_{ur} = -1 + \cO(\frac{1}{r^2}), \hspace{0,5 cm} g_{uz} = \cO(1) \ ,\\
& g_{zz} = \cO(r), \hspace{0,5 cm} g_{z\Bar{z}} = r^2\gamma_{z\Bar{z}}+\cO(1), \hspace{0,5 cm} g_{rr}=g_{rz}=0 \ .
\end{aligned}
\end{equation} 
Let's note some facts about the functions appearing in this expansion:
\begin{itemize}
    \item $m_{B}$ is called the Bondi mass aspect. For Kerr spacetimes $m_B=GM$ is constant, but generically it will depend on $u$ and $z$. Integrating it over the sphere gives the total Bondi mass.
    \item $N_z$ is called the angular momentum aspect since its integrals over the sphere, contracted with a rotational vector field, give the total angular momentum.
    \item $C_{zz}$ is called the shear and describes gravitational waves. We can use it to define the Bondi news tensor $N_{zz}=\partial_{u}C_{zz}$. It is transverse to $\cI^+$ and its square is proportional to the energy flux across $\cI^+.$
\end{itemize}
Lastly, we impose an additional boundary condition on $N_{zz}$, in particular that near $\cI^+_+$ it falls off faster than $\frac{1}{|u|}$.  This makes $C_{zz}$ well defined at the boundaries $\cI^+_+$ and $\cI^+_-$ which will allow us to write down a similar matching condition to the one we saw in the spin 1 case. Note that in \cite{Christodoulou1989-1990} it was proven that if $N_{zz}$ falls off at least as fast as $\frac{1}{|u|^{3/2}}$ then metric solutions of the type we assume exist and are dynamically stable and geodesically complete. We, however, will only need the weaker condition that $N_{zz}$ falls off faster than $\frac{1}{|u|}$. Similarly to the above discussion, using the advanced Bondi coordinates defined in (\ref{advcoord}) one can show that near $\cI^-$ the metric is constrained to be
\begin{equation}
\begin{aligned}
ds^2 =& -dv^2 +2dvdr +2r^2\gamma_{\z\Bar{\z}}d\z d\Bar{\z}+\frac{2m_B}{r}dv^2+rC_{\z\z}d\z^2+rC_{\Bar{\z}\Bar{\z}}d\Bar{\z}^2\\&-D^{\z}C_{\z\z}dvd\z-D^{\Bar{\z}}C_{\Bar{\z}\Bar{\z}}dvd\Bar{\z}
+... \  ,
\end{aligned}
\end{equation}
 and we define the Bondi news as $N_{\z\z} = \partial_{v}C_{\z\z}$ with a similar falloff condition. Note that so far we have written down the most general asymptotic metric assuming our falloff conditions and coordinate gauge. The metric near $\cI^+$ then depends on the values of $m_B$, $N_z$ and $C_{zz}$. To find these we need to integrate the Einstein equations, and we need initial data. Specifically, we need to know $N_{zz}$ on $\cI^+$, from which we can determine $m_B$ and $C_{zz}$ on $\cI^+$ up to integration functions. Thus, we can use the Einstein equations to fully determine the metric up to this order in $r$ given the initial data
 \begin{equation}
     \{N_{zz}(u,z,\bar{z}), C(z,\bar{z})|_{\cI^+_-},m_B(z,\bar{z})|_{\cI^+_-}\} \ .
 \end{equation}
 Here $C(z,\bar{z})|_{\cI^+_-}$ is the integration function used to fully determine $C_{zz}$. It is defined by 
 \begin{equation}
     C_{zz}|_{\cI^+_-} = -2D^2_{z}C|_{\cI^+_-} \ .
 \end{equation}
The statement that we have the freedom to choose $C$ and $m_B$ at $\cI^+_-$ is what we mean when we say the Einstein-Hilbert equations do not have a unique solution. It would be nice if this freedom could be fixed using knowledge of the metric near $\cI^-$.\footnote{Obviously, the metric near $\cI^-$ is also determined by integration using similar initial conditions at $\cI^-_+.$} How could we do this? The answer, as proposed in \cite{Strominger_2014}, is that we must match the initial conditions at $\cI^+_-$ and $\cI^-_+$ using 
\begin{equation}
\begin{aligned}
\label{matching}
z=\z \ \implies \ C(z,\Bar{z})|_{\cI^+_-} =& \  C(\z,\Bar{\z})|_{\cI^-_+} \ ,\\
z=\z \ \implies \ m_B(z,\Bar{z})|_{\cI^+_-}=& \ m_B(\z,\Bar{\z})|_{\cI^-_+} \ .
\end{aligned}
\end{equation}
Since $\z$ is defined in an antipodal sense compared to $z$ this matching condition is often stated by saying that $C|_{\cI^+_-}$ is antipodally related to $\cI^-_+$. As in electromagnetism, this matching condition implies the conservation of asymptotic charges at any point on the celestial sphere.

\subsubsection{Defining the charges}
Assuming our matching conditions in (\ref{matching}), given a function $\varepsilon:S^2\rightarrow \R$ we can immediately write down the following classical asymptotically conserved charges\footnote{As in the spin 1 case we will be interested in the choice $\varepsilon(z,\bar{z}) = \gamma^{-1}_{z\bar{z}}\delta^2(z-w)$ so $\varepsilon$ is best described as a distribution on $S^2$. 
}
\begin{equation}
\begin{aligned}
&Q^+_{\varepsilon} = \frac{1}{4\pi G}\int_{\cI^+_-}d^2z \gamma_{z\Bar{z}}\varepsilon(z,\bar{z})m_B(z,\bar{z}) \ , \\
&Q^-_{\varepsilon} = \frac{1}{4\pi G}\int_{\cI^-_+}d^2\z \gamma_{\z\Bar{\z}}\varepsilon(\z,\bar{\z})m_B(\z,\bar{\z}) \ .
\end{aligned}
\end{equation}
Using the leading form of the $uu$ component of the Einstein-Hilbert equations given by 
\begin{equation}
\partial_u m_B = \frac{1}{4}(D^2_zN^{zz}+D^2_{\Bar{z}}N^{\Bar{z}\Bar{z}})- T_{uu}, \quad T_{uu} \equiv \frac{1}{4}N_{zz}N^{zz} \ ,
\end{equation}
and integration by parts we can write these as
\begin{equation}
\begin{aligned}
Q^+_{\varepsilon} = \frac{1}{4\pi G}\int_{\cI^+} dud^2z \gamma_{z\Bar{z}}\bigg(&\frac{1}{4}\varepsilon N_{zz}N^{zz}-\frac{1}{4}(D^2_z\varepsilon N^{zz}+D^2_{\Bar{z}}\varepsilon N^{\Bar{z}\Bar{z}})\bigg)\\
&-4\pi G\int_{i^+} d\rho d^2z\frac{\rho^2 \gamma_{z\bar{z}}}{\sqrt{1+\rho^2}}\varepsilon_{H}T^M_{\tau\tau} \ , \\
Q^-_{\varepsilon} = \frac{1}{4\pi G}\int_{\cI^-} dvd^2\z \gamma_{\z\Bar{\z}}\bigg(&\frac{1}{4}\varepsilon N_{zz}N^{zz}+\frac{1}{4}(D^2_{\z}\varepsilon N^{\z\z}+D^2_{\Bar{z}}\varepsilon N^{\Bar{\z}\Bar{\z}})\bigg)\\
&-4\pi G\int_{i^-} d\rho d^2\z\frac{\rho^2 \gamma_{\z\bar{\z}}}{\sqrt{1+\rho^2}}\varepsilon_{H}T^M_{\tau\tau} \ .
\end{aligned}
\end{equation}
Here we have defined
\begin{equation}
\label{fHdef}
    \begin{aligned}
        &\varepsilon_{H}(\rho,z,\bar{z})=-\int \frac{d^2w}{4\pi}\frac{\varepsilon(w,\bar{w})\g_{w\bar{w}}}{(-\sqrt{1+\rho^2}+\rho \hat{\bx}_{z}\cdot \hat{\bx}_{w})^3} \ .
    \end{aligned}
\end{equation}
Next our goal is to find the form of these charges in the quantum theory and prove their asymptotic conservation.

\subsection{Quantum asymptotic charges as detectors}
Our convention for the mode expansion is given by
\begin{equation}
    \begin{aligned}
        h_{\mu\nu}(x) =\int_{p_0>0}\frac{d^4p}{(2\pi)^{3}}\delta(p^2)\sum_{r=\pm}(\a_r(p)\e^{*r}_{\mu\nu}(p)e^{ipx}+\a_{r}^{\dagger}(p)\e^{r}_{\mu\nu}(p)e^{-ipx}) \ .
    \end{aligned}
\end{equation}
In our retarded Bondi coordinates $(u,r,z,\bar{z})$, given a null momentum  $p=(0,1,z,\bar{z})$ we choose spin 1 polarization vectors according to
\begin{equation}
    \begin{aligned}
        \e^{1, \mu}(p) = \frac{1}{\sqrt{2}}(\bar{z},1,-i,-\bar{z}), \quad \e^{2, \mu}(p) = \frac{1}{\sqrt{2}}(z,1,i,-z) \ .
    \end{aligned}
\end{equation}
Note that of course the above are vectors defined at $x$. If we take $x=r p$, then we have
\begin{equation}
\begin{aligned}
    &\e^{1}_{z}\bigg|_{rp}(p) = \frac{\partial x_{\mu}}{\partial z}\bigg|_{rp}\e^{+, \mu}(p)=\frac{r}{\sqrt{2}}(\frac{(1-\bar{z}^2)}{(1+z\bar{z})^2}-\frac{(1+\bar{z}^2)}{(1+z\bar{z})^2}+\frac{2\bar{z}^2}{(1+z\bar{z})^2})= 0 \ ,\\
    &
\e^1_{\bar{z}}\bigg|_{rp}(p)= \frac{\partial x_{\mu}}{\partial \bar{z}}\bigg|_{rp}\e^{+, \mu}(p) = \frac{1}{\sqrt{2}}(\frac{(1-z^2)}{(1+z\bar{z})^2}+\frac{(1+z^2)}{(1+z\bar{z})^2}+\frac{2z\bar{z}}{(1+z\bar{z})^2})=\frac{r\sqrt{2}}{1+z\bar{z}} \ .
\end{aligned}
\end{equation}
Similarly, $\e^2_z\bigg|_{rp}(p) =\frac{r\sqrt{2}}{1+z\bar{z}}$ and $\e^2_{\bar{z}}\bigg|_{rp}(p)=0$. Also, note that $\e^1_u\bigg|_{rp}(p) = -\frac{\bar{z}}{\sqrt{2}}$ and $\e^2_u\bigg|_{rp}(p) = -\frac{z}{\sqrt{2}}$. We can form our spin 2 basis using
\begin{equation}
    \e^{i}_{\mu\nu}\bigg|_{x}(p) = \e^{i}_{\mu}\bigg|_{x}(p)\e^{i}_{\nu}\bigg|_{x}(p) \ .
\end{equation}
According to the above, we therefore have
\begin{equation}
    \begin{aligned}
        \e^1_{zz}\bigg|_{rp}(p)=\e^1_{z\bar{z}}\bigg|_{rp}(p)=\e^2_{z\bar{z}}\bigg|_{rp}(p)=\e^2_{\bar{z}\bar{z}}\bigg|_{rp}(p)=0, \quad \e^1_{\bar{z}\bar{z}}\bigg|_{rp}(p)=e^2_{zz}\bigg|_{rp}(p) = \frac{2r^2}{(1+|z|^2)^2} \ , 
    \end{aligned}
\end{equation}
and
\begin{equation}
    \e^1_{uu}\bigg|_{rp}(p)=\frac{\bar{z}^2}{2}, \quad \e^2_{uu}\bigg|_{rp}(p)=\frac{z^2}{2} \ .
\end{equation}
From these we can obtain
\begin{equation}
    \e^{1\bar{z}\bar{z}}\bigg|_{rp}(p)=\e^{1z\bar{z}}\bigg|_{rp}(p)=\e^{2z\bar{z}}\bigg|_{rp}(p)=\e^{2zz}\bigg|_{rp}(p)=0, \quad \e^{1zz}\bigg|_{rp}(p)=e^{2\bar{z}\bar{z}}\bigg|_{rp}(p) = \frac{(1+|z|^2)^2}{2r^2} \ .
\end{equation}
Before we can rewrite our charges in terms of detectors we need to do some work. We take dynamics to be governed by the Einstein equations 
\begin{equation}
    R_{\mu\nu} - \frac{1}{2}g_{\mu\nu}R = 8\pi G T^M_{\mu\nu} \ ,
\end{equation}
Note that our choice of metric implies
\begin{equation}
    C_{zz} = \frac{2\sqrt{8\pi G}}{r}h_{zz} \ .
\end{equation}
Thus we can write 
\begin{equation}
\begin{aligned}
    Q^+_{\varepsilon} = \frac{1}{4\pi G}\int_{\cI^+}du d^2z \gamma_{z\bar{z}}\bigg(&\varepsilon\frac{32 \pi G}{4r^2}\partial_u h_{zz}\partial_u h^{zz}-\frac{2\sqrt{8\pi G}}{4r}(D^2_z\varepsilon \ \partial_u h^{zz}+D^2_{\bar{z}}\varepsilon \ \partial_u h^{\bar{z}\bar{z}})\bigg)\\
    &-4\pi G\int_{i^+} d\rho d^2z\frac{\rho^2 \gamma_{z\bar{z}}}{\sqrt{1+\rho^2}}\varepsilon_{H}T^M_{\tau\tau} \ .
\end{aligned}
\end{equation}
We chose $T^M_{\mu\nu}$ to be sourced by a massive scalar field $\phi$ with expansion
\begin{equation}
    \phi(x) = \int \frac{d^3p}{(2\pi)^3}\frac{1}{2\w_p}(e^{i p x}a(p) +e^{-ipx}a^{\dagger}(p)) \ ,
\end{equation}
so we have
\begin{equation}
    T^M_{\mu\nu} = \partial_{\mu}\phi\partial_{\nu}\phi -\frac{1}{2} \eta_{\mu\nu}(\partial_{\rho}\phi \partial^{\rho}\phi+m^2\phi^2) \ .
\end{equation}
These are classical statements, and we need to write down the appropriate operator that we will use in the quantum theory. To do this we will use similar definitions to those discussed in the spin 1 case. We can put a massless operator $X$ at future null infinity by defining
\begin{equation}
    X(u,y) = \lim_{L\rightarrow \oo}L^{t_{X}}X(x+Ly) \ ,
\end{equation}
where $y=(u=0,r=1,z,\bar{z})$ is a normalised, future pointing null vector, $u = -y\cdot x$, and $t_{X}$ is chosen to ensure the limit is finite. We also define
\begin{equation}
    L_{\w_0}[X](\oo,y )=\frac{1}{2}\int^{\oo}_{-\oo} du(e^{-i\w_0 u}+e^{i\w_0 u})X(u,y) \ .
\end{equation}
Again, we also define a massive detector
\begin{equation}
    T^M_{\tau \tau}(\rho,y)=\lim_{\tau \rightarrow \oo}\tau^3T^M_{\tau\tau}(x) \ .
\end{equation}
Thus we can write
\begin{equation}
\begin{aligned}
    Q^+_{\varepsilon} =& \int d^2z \gamma_{z\bar{z}}\frac{1}{2}\varepsilon L_{\w_0}[\partial_u h_{zz}\partial_u h^{zz}](\oo,y)-\int_{i^+} d\rho d^2z\frac{\rho^2 \gamma_{z\bar{z}}}{\sqrt{1+\rho^2}}\varepsilon_{H}T^M_{\tau\tau}(\rho,y)\\
    &-\int d^2z \gamma_{z\bar{z}}\frac{1}{\sqrt{8\pi G}}\bigg(D^2_z\varepsilon L_{\w_0}[\partial_u h^{zz}](\oo,y)+D^2_{\bar{z}}\varepsilon L_{\w_0}[\partial_u h^{\bar{z}\bar{z}}](\oo,y)\bigg) \ .
\end{aligned}
\end{equation}
Here there is an implicit limit $\w_0\rightarrow0^+$ which is always to be taken after all other limits. If we instead work in advanced Bondi coordinates near $\cI^-$ we obtain 
\begin{equation}
\begin{aligned}
    Q^-_{\varepsilon} = \int d^2\z \gamma_{\z\bar{\z}}&\frac{1}{2}\varepsilon L_{\w_0}[\partial_v h_{\z\z}\partial_v h^{\z\z}](-\oo,y)-\int_{i^-} d\rho d^2\z\frac{\rho^2 \gamma_{\z\bar{\z}}}{\sqrt{1+\rho^2}}\varepsilon_{H}T^M_{\tau\tau}(\rho,y)\\
    &+\int d^2z \gamma_{z\bar{z}}\frac{1}{\sqrt{8\pi G}}\bigg(D^2_\z \varepsilon L_{\w_0}[\partial_v h^{\z\z}](-\oo,y)+D^2_{\bar{\z}}\varepsilon L_{\w_0}[\partial_v h^{\bar{\z}\bar{\z}}](-\oo,y)\bigg) \ .
\end{aligned}
\end{equation}
where now $y=(v=0,r=1,\z,\bar{\z})$ is past pointing. Of course, it will be easier to work with these charges if we take the stationary phase approximation. Doing so for the terms appearing in $Q^+_{\varepsilon}$ yields
\begin{equation}
\label{statphaseapprox1}
    \begin{aligned}
        &L_{\w_0}[\partial_u h_{zz}\partial_u h^{zz}](\oo,y)=\int_{0}^{\oo}d\b\frac{\b^2}{4(2\pi)^3}(\a^{\dagger}_{1,\cI^+}(\b y)\a_{1,\cI^+}(\b y)+\a^{\dagger}_{2,\cI^+}(\b y)\a_{2,\cI^+}(\b y)) \ ,\\
        &L_{\w_0}[\partial_{u}h^{zz}](\oo,y) = -\frac{|\w_0|(1+|z|^2)^2}{16\pi}(\a^{\dagger}_{1,\cI^+}(|\w_0| y)+\a_{2,\cI^+}(|\w_0| y)) \ ,\\
        &L_{\w_0}[\partial_{u}h^{\bar{z}\bar{z}}](\oo,y) =  -\frac{|\w_0|(1+|z|^2)^2}{16\pi}( \a^{\dagger}_{2,\cI^+}(|\w_0| y)+\a_{1,\cI^+}(|\w_0| y)) \ ,\\
        &T^M_{\tau\tau}(\rho,y)=\frac{m^3}{2(2\pi)^3}a_{i^+}^{\dagger}(\rho m\hat{\b}) a_{i^+}(\rho m\hat{\b}) \ .
    \end{aligned}
\end{equation}
Naturally, similar results follow for the terms in $Q^-_{\varepsilon}$. Now we can compute the following commutators with the scalar creation and annihilation operators:
\begin{equation}
    \begin{aligned}
        &[Q^+_{\varepsilon},a_{i^+}(p)] =- m\varepsilon_H(\rho_p,z_p,\bar{z}_p)a_{i^+}(p) \ ,\\
        &[Q^+_{\varepsilon},a^{\dagger}_{i^+}(p)] = +m\varepsilon_H(\rho_p,z_p,\bar{z}_p)a^{\dagger}_{i^+}(p) \ ,\\
        &[Q^-_{\varepsilon},a_{i^-}(p)] =- m\varepsilon_H(\rho_p,z_p,\bar{z}_p)a_{i^-}(p) \ ,\\
        &[Q^-_{\varepsilon},a^{\dagger}_{i^-}(p)] = +m\varepsilon_H(\rho_p,z_p,\bar{z}_p)a^{\dagger}_{i^-}(p) \ .\\
    \end{aligned}
\end{equation}
We also compute commutators with the graviton creation and annihilation operators:
\begin{equation}
    \begin{aligned}
        &[Q^+_{\varepsilon},\a_{i,\cI^+}(p)] = -\w_p\varepsilon(z_p,\bar{z}_p)\a_{i,\cI^+}(p) \ ,\\
        &[Q^+_{\varepsilon},\a^{\dagger}_{i,\cI^+}(p)] = +\w_p\varepsilon(z_p,\bar{z}_p)\a^{\dagger}_{i,\cI^+}(p) \ ,\\
        &[Q^-_{\varepsilon},\a_{i,\cI^-}(p)] = -\w_p\varepsilon(z_p,\bar{z}_p)\a_{i,\cI^-}(p) \ ,\\
        &[Q^-_{\varepsilon},\a^{\dagger}_{i,\cI^-}(p)] = +\w_p\varepsilon(z_p,\bar{z}_p)\a^{\dagger}_{i,\cI^-}(p) \ .\\
    \end{aligned}
\end{equation}
We are now ready to prove asymptotic conservation.
\subsection{Asymptotic conservation for undressed external states}
Now we will show that for arbitrary scattering states $\bra{\mathrm{out}}$ and $\ket{\mathrm{in}}$ without a Faddeev-Kulish dressing we have 
\begin{equation}
\label{undressed1}
    \bra{\mathrm{out}}Q^+_{\varepsilon}\ket{\mathrm{in}}-\bra{\mathrm{out}}Q^-_{\varepsilon}\ket{\mathrm{in}}=0 \ .
\end{equation}
This was first shown for the case we consider (a massive scalar coupled to gravity) in \cite{Campiglia_2015}. Thus let the $\ket{\mathrm{in}}$ state consists of $n_s$ scalar particles and $n_g$ gravitons, and the $\bra{\mathrm{out}}$ state of $n'_s$ particles and $n'_g$ gravitons, with momenta and energies indexed by $s\in\{1,...,n_s\}$, $g\in\{1,...,n_g\}$, $s'\in\{1,...,n'_s\}$, and $g'\in\{1,...,n'_g\}$ respectively. As we did in the spin 1 case we split the charge $Q^{\pm}_{f}$ into soft and hard parts, where the soft part creates soft gravitons and the hard part counts external hard particles with a weighting
\begin{equation}
    Q^{+}_{\varepsilon}=Q^{+}_{\varepsilon, \ \mathrm{soft}}+Q^{+}_{\varepsilon, \ \mathrm{hard}} \ ,
\end{equation}
where 
\begin{equation}
    \begin{aligned}
        &Q^{+}_{\varepsilon, \ \mathrm{soft}} = -\frac{1}{\sqrt{8\pi G}}\int d^2z \gamma_{z\bar{z}}\bigg((D^2_z\varepsilon L_{\w_0}[\partial_u h^{zz}](\oo,y)+D^2_{\bar{z}}\varepsilon L_{\w_0}[\partial_u h^{\bar{z}\bar{z}}](\oo,y)\bigg) \ ,\\
        &Q^{+}_{\varepsilon, \ \mathrm{hard}}=\int d^2z \gamma_{z\bar{z}}\frac{1}{2}\varepsilon L_{\w_0}[\partial_u h_{zz}\partial_u h^{zz}](\oo,y)-\int_{i^+} d\rho d^2z\frac{\rho^2 \gamma_{z\bar{z}}}{\sqrt{1+\rho^2}}\varepsilon_{H}T^M_{\tau\tau}(\rho,y) \ .
    \end{aligned}
\end{equation}
Of course, a similar separation applies for $Q^-_{\varepsilon}$. The contribution of $Q^{\pm}_{\varepsilon, \ \mathrm{hard}}$ to the left hand side of (\ref{undressed1}) is
\begin{equation}
\label{hardchargeundressedcont}
\begin{aligned}
    &\bra{\mathrm{out}}Q^+_{\varepsilon, \ \mathrm{hard}}\ket{\mathrm{in}} - \bra{\mathrm{out}}Q^-_{\varepsilon, \ \mathrm{hard}}\ket{\mathrm{in}}\\
    =&\bra{\mathrm{out}}\ket{\mathrm{in}}\bigg(\sum_{s'=1}^{n'_s}m\varepsilon_H(\frac{|\bp_{s'}|}{m},z_{s'},\bar{z}_{s'})-\sum_{s=1}^{n_s}m\varepsilon_H(\frac{|\bp_{s}|}{m},z_{s},\bar{z}_{s})\\&+\sum_{g'=1}^{n'_g}\w_{g'}\varepsilon(z_{g'},\bar{z}_{g'})-\sum_{g=1}^{n_g}\w_{g}\varepsilon(z_{g},\bar{z}_{g})\bigg) \ .
\end{aligned}
\end{equation}
Furthermore, we saw in (\ref{statphaseapprox1}) that the stationary phase approximation implies that $Q^{\pm}_{\varepsilon, \ \mathrm{soft}}$ acts to create soft gravitons. Specifically, using the soft theorem and integrating by parts twice, we find 
\begin{equation}
    \begin{aligned}
        &\bra{\mathrm{out}}Q^+_{\varepsilon, \ \mathrm{soft}}\ket{\mathrm{in}} 
    =\bra{\mathrm{out}}\ket{\mathrm{in}}\bigg(\sum^{n'_g}_{g'=1}\int d^2z \g_{z\bar{z}}\varepsilon(z,\bar{z})D^2_{\bar{z}}\frac{p^{\mu}_{g'}p^{\nu}_{g'}\e^1_{\mu\nu}}{8\pi \g_{z\bar{z}}p_{g'}\cdot y}\\
        &-\sum^{n_g}_{g=1}\int d^2z \g_{z\bar{z}}\varepsilon(z,\bar{z})D^2_{\bar{z}}\frac{p^{\mu}_{g}p^{\nu}_{g}\e^{1}_{\mu\nu}}{8\pi \g_{z\bar{z}}p_{g}\cdot y}
        +\sum^{n'_g}_{g'=1}\int d^2z \g_{z\bar{z}}\varepsilon(z,\bar{z})D^2_{z}\frac{p^{\mu}_{g'}p^{\nu}_{g'}\e^2_{\mu\nu}}{8\pi \g_{z\bar{z}}p_{g'}\cdot y}\\
        &-\sum^{n_g}_{g=1}\int d^2z \g_{z\bar{z}}\varepsilon(z,\bar{z})D^2_{z}\frac{p^{\mu}_{g}p^{\nu}_{g}\e^{2}_{\mu\nu}}{8\pi \g_{z\bar{z}}p_{g}\cdot y}
    +\sum^{n'_s}_{s'=1}\int d^2z \g_{z\bar{z}}\varepsilon(z,\bar{z})D^2_{\bar{z}}\frac{p^{\mu}_{s'}p^{\nu}_{s'}\e^1_{\mu\nu}}{8\pi \g_{z\bar{z}}p_{s'}\cdot y}\\
        &-\sum^{n_s}_{s=1}\int d^2z \g_{z\bar{z}}\varepsilon(z,\bar{z})D^2_{\bar{z}}\frac{p^{\mu}_{s}p^{\nu}_{s}\e^{1}_{\mu\nu}}{8\pi \g_{z\bar{z}}p_{s}\cdot y}
        +\sum^{n'_s}_{s'=1}\int d^2z \g_{z\bar{z}}\varepsilon(z,\bar{z})D^2_{z}\frac{p^{\mu}_{s'}p^{\nu}_{s'}\e^2_{\mu\nu}}{8\pi \g_{z\bar{z}}p_{s'}\cdot y}\\
        &-\sum^{n_s}_{s=1}\int d^2z \g_{z\bar{z}}\varepsilon(z,\bar{z})D^2_{z}\frac{p^{\mu}_{s}p^{\nu}_{s}\e^{2}_{\mu\nu}}{8\pi \g_{z\bar{z}}p_{s}\cdot y}\bigg) \ .
    \end{aligned}
\end{equation}
Note that the graviton momenta $p_g$ are massless, and the scalar momenta $p_s$ are massive. We find
\begin{equation}
    D^2_{z}\frac{p^{\mu}_{g}p^{\nu}_{g}\e^{2}_{\mu\nu}}{ \g_{z\bar{z}}p_{g}\cdot y}=D^2_{\bar{z}}\frac{p^{\mu}_{g}p^{\nu}_{g}\e^{1}_{\mu\nu}}{ \g_{z\bar{z}}p_{g}\cdot y} = -\frac{2\pi\w_g}{\gamma_{z\bar{z}}}\delta^2(z-z_g) \ ,
\end{equation}
and
\begin{equation}
    D^2_{z}\frac{p^{\mu}_{s}p^{\nu}_{s}\e^{2}_{\mu\nu}}{ \g_{z\bar{z}}p_{s}\cdot y}=D^2_{\bar{z}}\frac{p^{\mu}_{s}p^{\nu}_{s}\e^{1}_{\mu\nu}}{ \g_{z\bar{z}}p_{s}\cdot y} = \frac{m}{2(-\sqrt{1+\rho^2_{s}}+\rho_{s}\hat{\bx}_{s}\cdot \hat{\bx}_{z})^3} \ .
\end{equation}
Clearly, similar results follow for $Q^{-}_{\varepsilon, \ \mathrm{soft}}$. Thus, recalling our definition of $\varepsilon_{H}$ in (\ref{fHdef}), and substituting everything in, we immediately find the required result
\begin{equation}
    \bra{\mathrm{out}}Q^+_{\varepsilon}\ket{\mathrm{in}}-\bra{\mathrm{out}}Q^-_{\varepsilon}\ket{\mathrm{in}}=0 \ .
\end{equation}
As in the case with QED, this result is equivalent to the soft theorem. Thus we have used detectors to confirm the result in \cite{Campiglia_2015}. 
\subsection{Asymptotic conservation for dressed external states}
In the previous section, our proof of conservation relied on the graviton soft theorem, which introduces poles which cancel with soft factors in $Q^{\pm}_{\varepsilon, \ \mathrm{soft}}$. We previously used the fact that, in massive QED, one can instead construct `physical' scattering states which remove IR divergences from S-matrix calculations. Similar dressings have been constructed for gravity \cite{Ware_2013}. Note that because we must also dress external gravitons, there are remaining collinear singularities, but these are square integrable \cite{Prabhu:2022zcr} and so do not affect the well defined-ness of our distribution valued operators. Now we will show that when we consider these FK dressed scattering states the conservation of $Q_{\varepsilon}$ still holds
\begin{equation}
    \begin{aligned}
        &\bra{\mathrm{out}, \  \mathrm{dressed}}Q^+_{\varepsilon}\ket{\mathrm{in}, \ \mathrm{dressed}}-\bra{\mathrm{out}, \  \mathrm{dressed}}Q^-_{\varepsilon}\ket{\mathrm{in}, \ \mathrm{dressed}}=0 \ .
    \end{aligned}
\end{equation}
as was first shown to hold for no hard external gravitons in \cite{Choi_2018}. Here we will extend this to any scattering process by including the full $t$ dependence of the FK dressings. We will also find that the contribution to the memory effect from the FK dressing is non-zero. This is unlike \cite{Choi_2018} which incorrectly determines it to be zero.  

Let's thus discuss how to define dressed scattering states for the case of a massive scalar coupled to gravity. Again, let the $\ket{\mathrm{in}}$ state consists of $n_s$ scalar particles and $n_g$ gravitons, and the $\bra{\mathrm{out}}$ state of $n'_s$ particles and $n'_g$ gravitons, with momenta and energies indexed by $s\in\{1,...,n_s\}$, $g\in\{1,...,n_g\}$, $s'\in\{1,...,n'_s\}$, and $g'\in\{1,...,n'_g\}$ respectively. Then we define
\begin{equation}
    \begin{aligned}
        &\ket{\mathrm{in}, \ \mathrm{dressed}} = \prod_{s=1}^{n_s}\prod_{g=1}^{n_g}e^{\Phi(p_s)}e^{\Phi(p_g)}\ket{\mathrm{in}} \ ,\\
        &\bra{\mathrm{out}, \ \mathrm{dressed}} = \bra{\mathrm{out}}\prod_{s'=1}^{n'_s}\prod_{g'=1}^{n'_{g}}e^{-\Phi(p_{g'})}e^{-\Phi(p_{s'})} \ ,
    \end{aligned}
\end{equation}
where
\begin{equation}
    \Phi(p) = \sqrt{8\pi G}\int\frac{d^3\bk}{(2\pi)^3}\frac{1}{2\w_k}[f(k,p)\cdot \e^{r}\a^{\dagger}_{r}(k)-f^*(k,p)\cdot\e^{r*}(k)\a_{r}(k)] \ .
\end{equation}
Here
\begin{equation}
    f_{\mu\nu}(k,p) = (\frac{p_{\mu}p_{\nu}}{k\cdot p}-c_{\mu\nu}(k,p))\varphi(k,p)e^{i2\pi \w_k t}\ ,
\end{equation}
where $\varphi(k,p)$ is a smoothing function satisfying $\varphi(k,p)=1$ in a neighbourhood of $k=0$ and $\lim_{\w_k\rightarrow \oo}\varphi(k,p)=0$ where the decay is exponential, and $c_{\mu\nu}$ satisfies $c_{\mu\nu}=c_{\nu\mu}$ , $c_{\mu\nu}k^{\mu}=p_{\nu}$ , and $c^{\mu\nu}c_{\mu\nu}-(c^{\mu}{}_{\mu})^2/2=0$ . We agree to take the limit $t\rightarrow +\oo$ at the end of any S-matrix calculation.

This dressing without the exponential was constructed and shown to cancel all IR divergences in \cite{Ware_2013}. The extra exponential arises in the derivation of the dressing and cannot be dropped if one includes external gravitons, since it represents the $t$ dependence of the dressing and gravitons self interact.\footnote{Here $t$ is related to the time $t'$ which is used in e.g. \cite{Kulish:1970ut} by $t=\frac{p\cdot \hat{k}}{2\pi \w_p}t'$. Strictly speaking the time $t'$ depends on the momentum of the particle that one is dressing as in \cite{Kulish:1970ut}. Here we have performed a transformation for each particle such that the dressing can be written in the same way. Limits can be taken independently for each particle, so this is okay.} Note that in \cite{Ware_2013} the authors impose an additional constraint on $c_{\mu\nu}$ which we will not need. This constraint is used to construct Chung states which make it easy to show divergences cancel in explicit calculations at the cost that the dressing is not valid for the case of external gravitons \cite{Kulish:1970ut}. Instead, we note that we must use the unique rotationally invariant choice from \cite{Choi_2018}
\begin{equation}
    c_{\mu\nu} = \frac{q_{\mu}p_{\nu}+q_{\nu}p_{\mu}}{q\cdot k}-\frac{k\cdot p}{(q\cdot k)^2}q_{\mu}q_{\nu} \ .
\end{equation}
and we will show that it in fact does contribute physically to the memory effect unlike the result in \cite{Choi_2018}. Here  $q(k)$ is any map from null vectors $k$ to a null vector $q(k)\neq k$, such that $q(k)$ is homotopic to the antipodal map. We will choose $q=(1,-\bk)$, and it is not hard to show that this gives the unique rotationally invariant choice of $c_{\mu\nu}$. Let us review what we have done here. Each incoming and outgoing particle is dressed with an exponential factor which depends on the respective particle's momentum (note that the $p_{g'}$ and $p_g$ are null but the $p_{s'}$ and $p_s$ are massive). As we have noted, these dressings ensure that all S-matrix elements are IR finite \cite{Ware_2013}. Now let us show that using these dressed states gives
\begin{equation}
    \begin{aligned}
        &\bra{\mathrm{out}, \  \mathrm{dressed}}Q^+_{\varepsilon}\ket{\mathrm{in}, \ \mathrm{dressed}}-\bra{\mathrm{out}, \  \mathrm{dressed}}Q^-_{\varepsilon}\ket{\mathrm{in}, \ \mathrm{dressed}}=0 \ .
    \end{aligned}
\end{equation}
Before we do this however, we need to discuss some formal requirements on the limits involved in this calculation. The two limits of interest are the ${t\rightarrow\oo}$ limit which we have formally included in the definition of our dressings and the $\w_0\rightarrow 0^+$ limit which defines our distribution valued operators. Respectively, these limits can be physically interpreted as taking the dressings to be asymptotic as required and taking the photons in the operators to be soft. We now agree to define $\w_0 = 1/t$ and take these limits at the same time. This is not some convenient trick, in fact it is physically required for multiple reasons. For one, in $d=4$ it is known that both radiation and the memory effect decay as $1/r$. Thus the contribution due to the dressing, which describes memory, should falloff at the same rate as the energy of soft photons. The only correct choice on dimensional grounds is $\w_0t=\mathrm{constant}$. The exact choice of constant depends on the $t$ dependent exponential used in the definition of $f_{\mu\nu}$ and is easily determined as the choice which ensures conservation and the correct memory effect. We will also see that this set up is required for the dressing to not change the overall energy of the state, which we naturally expect. Note that these points are also true for QED, but we ignored this subtlety in the spin 1 case since it did not cause issues. With these formal agreements in place we can move on to calculating $[Q^+_{\varepsilon},\Phi(p)]$. Recall that we defined
\begin{equation}
\begin{aligned}
    Q^+_{\varepsilon} =& \int d^2z \gamma_{z\bar{z}}\frac{1}{2}\varepsilon L_{\w_0}[\partial_u h_{zz}\partial_u h^{zz}](\oo,y)-\int_{i^+} d\rho d^2z\frac{\rho^2 \gamma_{z\bar{z}}}{\sqrt{1+\rho^2}}\varepsilon_{H}T^M_{\tau\tau}(\rho,y)\\
    &-\int d^2z \gamma_{z\bar{z}}\frac{1}{\sqrt{8\pi G}}\bigg(D^2_z\varepsilon L_{\w_0}[\partial_u h^{zz}](\oo,y)+D^2_{\bar{z}}\varepsilon L_{\w_0}[\partial_u h^{\bar{z}\bar{z}}](\oo,y)\bigg) \ .
\end{aligned}
\end{equation}
where the stationary phase approximation gives
\begin{equation}
    \begin{aligned}
        &L_{\w_0}[\partial_u h_{zz}\partial_u h^{zz}](\oo,y)=\int_{0}^{\oo}d\b\frac{\b^2}{4(2\pi)^3}(\a^{\dagger}_{1,\cI^+}(\b y)\a_{1,\cI^+}(\b y)+\a^{\dagger}_{2,\cI^+}(\b y)\a_{2,\cI^+}(\b y)) \ ,\\
        &L_{\w_0}[\partial_{u}h^{zz}](\oo,y) = -\frac{|\w_0|(1+|z|^2)^2}{16\pi}(\a^{\dagger}_{1,\cI^+}(|\w_0| y)+\a_{2,\cI^+}(|\w_0| y)) \ ,\\
        &L_{\w_0}[\partial_{u}h^{\bar{z}\bar{z}}](\oo,y) =  -\frac{|\w_0|(1+|z|^2)^2}{16\pi}( \a^{\dagger}_{2,\cI^+}(|\w_0| y)+\a_{1,\cI^+}(|\w_0| y)) \ ,\\
        &T^M_{\tau\tau}(\rho,y)=\frac{m^3}{2(2\pi)^3}a_{i^+}^{\dagger}(\rho m\hat{\b}) a_{i^+}(\rho m\hat{\b}) \ .
    \end{aligned}
\end{equation}
Clearly $T_{\tau\tau}^M(\rho,y)$ commutes with $Q^+_{\varepsilon}$ and annihilates the vacuum. For the other three terms we find
\begin{equation}
\label{commutatorsdressed}
    \begin{aligned}
        &[L_{\w_0}[\partial_u h_{zz}\partial_u h^{zz}](\oo,y),\Phi(p)]=0 \ , \\
        &[L_{\w_0}[\partial_{u}h^{zz}](\oo,y),\Phi(p)]=-\sqrt{8\pi G}\frac{|\w_0|(1+|z|^2)^2}{16\pi}(f(|\w_0| y,p)\cdot\e^2+f^*(|\w_0| y,p)\cdot \e^{*1})  \ ,\\
        &[L_{\w_0}[\partial_{u}h^{\bar{z}\bar{z}}](\oo,y),\Phi(p)]=-\sqrt{8\pi G}\frac{|\w_0|(1+|z|^2)^2}{16\pi}(f(|\w_0| y,p)\cdot\e^1+f^*(|\w_0| y,p)\cdot \e^{*2}) \ .
    \end{aligned}
\end{equation}
Let us first show why the first commutator vanishes. A calculation gives
\begin{equation}
    [L_{\w_0}[\partial_u h_{zz}\partial_u h^{zz}](\oo,y),\Phi(p)]=\sqrt{8\pi G}\int^{\oo}_{0}d\b \frac{\b^2}{4(2\pi)^3}[f(\b y,p)\cdot\e^r a^{\dagger}_r(\b y)+f^*(\b y,p)\cdot \e^{*r}a_r(\b y)] \ .
\end{equation}
Recall that here 
\begin{equation}
    f_{\mu\nu}(\b y, p ) = (\frac{p_{\mu}p_{\nu}}{\b y\cdot p}-c_{\mu\nu}(\b y,p))\varphi(k,p)e^{i2\pi \b  t} \ .
\end{equation}
By the Riemann-Lebesgue lemma an integral over $\b$ of this decays as $1/t$ and will thus vanish in the limit $t\rightarrow \oo$. Note that we physically require this commutator to be zero anyway, since otherwise the dressing would change the energy of the scattering state. Combining all three commutators we have found
\begin{equation}
    [Q^+_{\varepsilon},\Phi(p)] = \frac{1}{8\pi}\int d^2z \gamma_{z\bar{z}} \varepsilon(D^2_z(\gamma^{-1}_{z\bar{z}}f(|\w_0| y,p)\cdot\e^2+\gamma^{-1}_{z\bar{z}}f^*(|\w_0| y,p)\cdot \e^{*1})+c.c.) \ .
\end{equation}
Note that in this case
\begin{equation}
    f_{\mu\nu}(\w_0 y,p) = (\frac{p_{\mu}p_{\nu}}{k\cdot p}-c_{\mu\nu}(k,p))\varphi(k,p)e^{i2\pi} \ .
\end{equation}
Thus the result is now $t$ independent. This occurred because we used the physically required $\w_0=1/t$. Now, since we have
\begin{equation}
    D^2_{z}\frac{p^{\mu}_{g}p^{\nu}_{g}\e^{2}_{\mu\nu}}{ \g_{z\bar{z}}p_{g}\cdot y}=D^2_{\bar{z}}\frac{p^{\mu}_{g}p^{\nu}_{g}\e^{1}_{\mu\nu}}{ \g_{z\bar{z}}p_{g}\cdot y} = -\frac{2\pi\w_g}{\gamma_{z\bar{z}}}\delta^2(z-z_g) \ ,
\end{equation}
and
\begin{equation}
    D^2_{z}\frac{p^{\mu}_{s}p^{\nu}_{s}\e^{2}_{\mu\nu}}{ \g_{z\bar{z}}p_{s}\cdot y}=D^2_{\bar{z}}\frac{p^{\mu}_{s}p^{\nu}_{s}\e^{1}_{\mu\nu}}{ \g_{z\bar{z}}p_{s}\cdot y} = \frac{m}{2(-\sqrt{1+\rho^2_{s}}+\rho_{s}\hat{\bx}_{s}\cdot \hat{\bx}_{z})^3} \ ,
\end{equation}
we find
\begin{equation}
    \bra{\mathrm{out}, \ \mathrm{dressed}}Q^+_{\varepsilon, \ \mathrm{soft}} = (-\sum_{g'=1}^{n'_g}\w_{g'}\varepsilon(z_{g'},\bar{z}_{g'})-\sum_{s'=1}^{n'_s}m\varepsilon_H(\rho_{s'},z_{s'},\bar{z}_{s'})+\cdots)\bra{\mathrm{out}, \ \mathrm{dressed}} \ ,
\end{equation}
where the $+\cdots$ refers to the contribution from $c_{\mu\nu}$. The first two summations exactly cancel the contribution from $T_{\tau\tau}^M(\rho,y)$ that we found in (\ref{hardchargeundressedcont}). A similar calculation follows for $Q^-_{\varepsilon}$. Thus we have finally shown that we indeed have
\begin{equation}
    \begin{aligned}
        &\bra{\mathrm{out}, \  \mathrm{dressed}}Q^+_{\varepsilon}\ket{\mathrm{in}, \ \mathrm{dressed}}-\bra{\mathrm{out}, \  \mathrm{dressed}}Q^-_{\varepsilon}\ket{\mathrm{in}, \ \mathrm{dressed}}=0 \ .
    \end{aligned}
\end{equation}
It is again interesting to consider the eigenvalue of $\bra{\mathrm{out}, \  \mathrm{dressed}}$ when we act with $Q^+_f$, which will depend on $c_{\mu\nu}$. Choosing
\begin{equation}
    \varepsilon(z,\bar{z})=\frac{1}{\gamma_{z\bar{z}}}\delta(z-z_{\hat{y}}) \ ,
\end{equation}
where 
\begin{equation}
    \hat{y} = (1,\frac{z_y+\bar{z_y}}{1+|z_y|^2},i\frac{-z_y+\bar{z_y}}{1+|z_y|^2},\frac{1-|z_y|^2}{1+|z_y|^2}) \ 
\end{equation} 
and defining the corresponding charge as $Q^+_{\hat{y}}$ we find
\begin{equation}
    \bra{\mathrm{out}, \ \mathrm{dressed}}Q^+_{\hat{y}} = \bigg(\frac{1}{4\pi}\sum_{g'=1}^{n'_g}(4\w_{g'}+3p_{g'}\cdot \hat{y})+\frac{1}{4\pi}\sum_{s'=1}^{n'_s}(4\w_{s'}+3p_{s'}\cdot \hat{y})\bigg)\bra{\mathrm{out}. \ \mathrm{dressed}} \ .
\end{equation}
Note that 
\begin{equation}
    4\w_{g'}+3p_{g'}\cdot \hat{y}=\w_{g'}+3\bp_{g'}\cdot \hat{\by} \ .
\end{equation}
The first term on the right hand side exactly mirrors the result in QED, with the charge naturally replaced by energy. We also find a dipole contribution which was absent in the QED case. These terms are due to the unique rotationally invariant $c_{\mu\nu}$ which we chose, and we see that it does indeed contribute to the memory, unlike what is claimed in \cite{Choi_2018}. As in the spin 1 case, FK dressings allow us to construct a basis of the Fock space which is made up of eigenvectors of $Q^{^+}_{\hat{y}}$. In the next section we will again interpret this result in terms of physical asymptotic Fock spaces and the memory effect.

\subsection{Physical asymptotic Fock spaces and the memory effect}
In the previous two subsections we focused on defining our charges as operator valued distributions and proving that these charges are conserved between arbitrary scattering states, both for the dressed and undressed case. Let us now briefly discuss the physical interpretation of our results. We will again rely on the discussions in \cite{Prabhu:2022zcr}. 

The memory effect of gravity is similar to that of QED; it is the physical effect where the field differs at late retarded times compared to early retarded times at order $1/r$. Whilst the memory effect in QED can be interpreted as a net momentum kick of a charged test particle at late retarded times \cite{Bieri_2013}, in gravity it should instead be thought of as a net displacement of a massive particle at late retarded times \cite{1974AZh....51...30Z}. All physical effects should be encoded into a complete field theory description, but it is not trivial to encode the memory effect in gravity into the asymptotic Fock spaces. As in QED, this is because the memory effect and radiation both decay as $1/r$ \cite{Satishchandran_2019}, and so we must add soft gravitons to capture the memory effect, which leads to well known IR divergences due to the Weinberg soft theorem. Thus, the Weinberg soft theorem is a direct manifestation of the failure of the traditional asymptotic Fock spaces to capture the memory effect. The IR divergences are cured by including FK dressings, and we will see that these dressings simultaneously correctly encode the memory effect. 

First, recall that one of our assumptions was that the Bondi news tensor $N_{AB}$, where $A,B\in \{z,\bar{z}\}$, falls off faster than $\frac{1}{|u|}$. Using this we may define the gravitational memory as \cite{Strominger:2014pwa}
\begin{equation}
    \Delta^{\mathrm{GR}}_{AB}\equiv \frac{1}{2}\int^{\oo}_{-\oo}du N_{AB}= \frac{1}{2}(C_{AB}|_{u=+\oo}-C_{AB}|_{u=-\oo}) \ .
\end{equation}
We call $\Delta^{\mathrm{GR}}_{AB}$ the `gravitational memory out' and it depends on both the incoming state and the outgoing state. We can relate it to our previously defined charges using \cite{Strominger:2014pwa, Prabhu:2022zcr}
\begin{equation}
    -\int_{S^2} d\Omega \Delta^{\mathrm{GR}}_{AB} D^AD^B \varepsilon = 8\pi GQ^+_{\varepsilon, \ \mathrm{soft}} \ .
\end{equation}
To measure the memory in a particular direction given by a null vector vector $\hat{y}$ where
\begin{equation}
    \hat{y}=(1,\frac{z_y+\bar{z_y}}{1+|z_y|^2},i\frac{-z_y+\bar{z_y}}{1+|z_y|^2},\frac{1-|z_y|^2}{1+|z_y|^2}) \ ,
\end{equation}
we must choose
\begin{equation}
    \varepsilon(z,\bar{z})=\frac{1}{\gamma_{z\bar{z}}}\delta^2(z-z_y) \ .
\end{equation}
Let us denote the charge given by this choice as $Q^+_{\hat{y}}$. Then we have
\begin{equation}
       Q^+_{\hat{y}, \ \mathrm{soft}} \ =-\frac{1}{8\pi G}D^AD^B\Delta^{\mathrm{GR}}_{AB}(z_y,\hat{z}_y) \ .
\end{equation}
The FK dressed out states are eigenstates of $Q^+_{\hat{y}, \ \mathrm{soft}}$ and satisfy
\begin{equation}
    \bra{\mathrm{out}, \ \mathrm{dressed}}Q^+_{\hat{y}, \ \mathrm{soft}}= \lambda(\hat{y})\bra{\mathrm{out}, \ \mathrm{dressed}} \ ,
\end{equation}
where
\begin{equation}
    \lambda(\hat{y})=\frac{1}{4\pi}\sum_{g'=1}^{n'_g}(4\w_{g'}+3p_{g'}\cdot \hat{y}-\frac{4\pi\w_{g'}}{\gamma_{z\bar{z}}}\delta^2(z_{g'}-z_y))+\frac{1}{4\pi}\sum_{s'=1}^{n'_s}(4\w_{s'}+3p_{s'}\cdot \hat{y}+\frac{m^4}{(p_{s'}\cdot \hat{y})^3}) \ .
\end{equation}
Note that whilst the total memory depends on the outgoing and incoming states, we only expect this eigenvalue to include the contribution from the outgoing state. This is because the memory due to the outgoing state is not dependent on the memory due to the ingoing state. A consistent, physical Fock space should thus encode only the memory due to the outgoing state in the outgoing Fock space. We can see that $\lambda(\hat{y})$ is exactly this by considering \cite{Bieri_2014}. Here Bieri and Garfinkle find
\begin{equation}
    D^AD^B\Delta^{\mathrm{GR}}_{AB}= \bigg(P(\oo)-\sum_{i=0,1}[P(\oo)]_{[i]}\bigg)-\bigg(P(-\oo)-\sum_{i=0,1}[P(-\oo)]_{[i]}\bigg)+8\pi\bigg(F-\sum_{i=0,1}[F]_{[i]}\bigg) \ .
\end{equation}
Here $F$ is the outgoing energy per unit solid angle due to massless particles and $P(u)$ is the leading component in $1/r$ of the electric part of the Weyl tensor fully projected onto the radial direction at retarded time $u$. We have defined $[X]_{[i]}$ as the $\ell=i$ part of $X$. Note that $P(-\oo)$ depends on the incoming state and so we expect to find (up to a choice of units)
\begin{equation}
    -\lambda = \bigg(P(\oo)-\sum_{i=0,1}[P(\oo)]_{[i]}\bigg)+8\pi\bigg(F-\sum_{i=0,1}[F]_{[i]}\bigg) 
\end{equation}
Indeed, it is not hard to show that this is exactly true.\footnote{When this paper was submitted we were not completely clear about how the results match. Bieri and Garfinkle do not include energy due to gravitons in $F$ since they only work to first order in perturbation theory. However, we do, as already predicted long ago by Thorne \cite{PhysRevD.45.520}, and we expect that this captures higher order corrections to the memory.} In particular, note that the monopole and dipole terms in $\lambda$ correspond to the $\ell=0$ and $\ell=1$ parts that we subtract. Again, due to energy and momentum conservation we expect these to cancel when we calculate the full memory including the contribution due to the incoming state. However, we argue that they should be included in the eigenvalue since they arise from the $c_{\mu\nu}$ term in the dressing which is required to gauge fix the Fock space. Thus we do expect physical state vectors to contain this contribution to the eigenvalue. Note that these terms then cannot be `large gauge transformed away' as claimed in \cite{Choi_2018}.

Since the dressing diagonalizes the memory operator, it is encoding the memory effect entirely into the outgoing Fock space. This fact lends itself to the interpretation that the dressed Fock space is `physical' in some sense.

\section{Conclusion}
\label{conclusion}
There are infinitely many charges which are conserved during scattering processes in theories which couple a scalar to massless spin 1 and spin 2 gauge fields \cite{He_2014, He_2015, Campiglia:2015qka, Campiglia_2015}. Here we have expressed these charges as detectors in the case of a massive scalar and confirmed their conservation. In particular we have shown conservation for scattering using both the ordinary Fock space states and Faddeev-Kulish dressed states in both theories. The precise language of detectors, in particular when considered as distribution valued operators, allows us to extend the spin 2 conservation to the general case including hard external gravitons. We use the full Faddeev-Kulish dressing in both cases, as opposed to Chung states, and note that imposing the respective gauge conditions on the Fock spaces leads to contributions to the memory effect from the commutator of the memory operator with the dressing. For QED this is proportional to the total charge of the scattering process, and for gravity it is the sum of two terms which are respectively proportional to the total energy of the process, and a dipole contribution which is absent in the QED case. These terms are present in the formulas given by Bieri and Garfinkle in \cite{Bieri_2013, Bieri_2014} which confirms the dressing has been chosen correctly. To the best of our knowledge we have obtained multiple corrections and contributions to the literature which we list here
\begin{itemize}
    \item We saw that the dressing is consistent with using the commutators which are smaller by a factor of $1/2$ than the usual result in the literature,  \cite{He_2014, He_2015, Gabai_2016, strominger2018lecturesinfraredstructuregravity, Kapec_2017, Arkani_Hamed_2021}. For example, for the spin 1 case 
    \begin{equation}
    \begin{aligned}
        &[Q^+_{\e}, A_z(u,y)]=-\frac{1}{2}\partial_{z}\e(z_y,\bar{z}_y) \ ,\\
        &[Q^+_{\e}, A_{\bar{z}}(u,y)]=-\frac{1}{2}\partial_{\bar{z}}\e(z_y,\bar{z}_y) \ ,\\
    \end{aligned}
\end{equation}
give a consistent shift in the dressing shift and are obtained using the standard commutation relations in the bulk. These commutators do not affect the proof of conservation when using undressed scattering states, but we require them to show conservation if one uses FK dressed states. There is an analogous commutator in the spin 2 case with the same story. For a more in depth discussion of the tension between soft mode commutators, dressing and BMS frames see Appendix \ref{AppendixA}.  
    \item We have found that the full dressing, including the $t$ dependence and correct choice of gauge fixing function $c_{\mu\nu}$ (the unique rotationally invariant choice) must be used to properly describe the memory effect as found in \cite{Bieri_2013, Bieri_2014}. This is obtained by simply not dropping the $t$ dependence which occurs in the derivation of the dressings in \cite{Kulish:1970ut, Ware_2013}. This is required since gravitons self interact and the memory effect is described by the dressing at order $1/t$. Using this dressing can be thought of as extending the Chung states to the full Faddeev-Kulish dressed states.
    \item Following the previous point, one must take the limit of the energy $\w_0$ of soft gravitons to zero with $t$, i.e. $\w_0 \sim 1/t$. The exact proportionality constant can be determined physically as the retarded time that a graviton in the dressing sees an external particle, or as the unique choice which ensures conservation of our asymptotic charges. This appears to resolve the question stated in the conclusion of \cite{Campiglia_2015}.
    \item The above two points allow us to extend all results to include scattering with external hard gravitons. Careful consideration of the detectors as distribution valued operators and the Riemann-Lebesgue lemma is used to do this, and show that the dressings do not contribute to the total energy. 
    \item We find via explicit calculation that the $c_{\mu\nu}$ term required to gauge fix the dressed Fock space in gravity does indeed contribute to the memory eigenvalue, and that the term is the expected analogue of the QED case plus a dipole term which is absent in the QED case. This term is present in the classical memory derived by Bieri and Garfinkle \cite{Bieri_2014}. Note that we use the same $c_{\mu\nu}$ as \cite{Choi_2018}, but we find that it leads to a non-vanishing BMS charge. 
    Note that this contribution is physical and due to the scattering process and cannot be `large gauge transformed' away. This term arises from extending the Chung states used in \cite{Strominger:2014pwa} to the full gauge fixed FK states. 
    The conservation of BMS charge in the trivial vacuum is then due to charge conservation in the case of QED, or energy and momentum conservation in the case of gravity. Note that we do NOT find an additional contribution to the full memory effect, as these terms cancel when one considers the contribution from both the incoming and outgoing states. However, since the $c_{\mu\nu}$ term is required in the dressing to restrict to the physical Fock space, our claim is simply that this term cannot be dropped when considering the eigenvalue of the memory operator when it acts on dressed states.
    \item Similarly, we find that the memory due to the out state in the case of QED for non-zero total charge is calculable and has an extra term proportional to the total charge which arises from the $c_{\mu}$ term in the dressing. Scattering processes with non-zero charge were already considered in \cite{Gabai_2016}. 
    We reiterate that this term arises when using the full gauge fixed Faddeev-Kulish dressing and cannot be large gauge transformed away, and can be shown to lead to agreement with the classical memory effect found in \cite{Bieri_2013}. As before, the term cancels when calculating the full memory effect due to the incoming and outgoing states.
\end{itemize}

\acknowledgments

We thank Laurent Friedel and Barak Gabai for useful discussions. I.M. and B.O. are supported by the DOE Early Career Award DE-SC0025581 and the Sloan Foundation.

\appendix

\section{Tracing factors of $\frac{1}{2}$}

\label{AppendixA}
The goal of this appendix is two fold: first, to clarify our conventions for the brackets in the dressing manipulations above and second to contextualize this in regards to the soft brackets used elsewhere. We will stick to the $U(1)$ case here for concreteness, but an analogous story follows straightforwardly for the leading soft theorem in gravity as well.

In (\ref{QcommutatorA}) we have 
\begin{equation}\
\label{appendixeq1}
    \begin{aligned}
        &[Q^+_{\e}, A_z(u,y)]=-\frac{1}{2}\partial_{z}\e(z_y,\bar{z}_y) \ ,\\
        &[Q^+_{\e}, A_{\bar{z}}(u,y)]=-\frac{1}{2}\partial_{\bar{z}}\e(z_y,\bar{z}_y) \ ,\\
    \end{aligned}
\end{equation}
which is a factor of 2 smaller than that in, for example, \cite{He_2014, Gabai_2016, strominger2018lecturesinfraredstructuregravity, Kapec_2017, Arkani_Hamed_2021}. Independently, one requires  
\begin{equation}
\label{appendixeq2}
    \begin{aligned}
        &[Q^+_{\e, \ \mathrm{soft}},\Phi(p)] =i\e_H(\rho_p,z_{p},\bar{z}_{p})+\frac{i}{4\pi}\int d^2z \gamma_{z\bar{z}}\e(z,\bar{z}) \ ,\\
        &[Q^-_{\e, \ \mathrm{soft}},\Phi(p)] =i\e_H(\rho_p,z_{p},\bar{z}_{p})+\frac{i}{4\pi}\int d^2z \gamma_{z\bar{z}}\e(z,\bar{z}) \ ,
    \end{aligned}
\end{equation}
where $\Phi(p)$ is the dressing factor defined in~\eqref{FKphidefqed}, and $\varepsilon_H$ is the gauge parameter propagated to $i^\pm$ as in~\eqref{Fkdressingcommutatorqed}, 
in order for $Q$ to be conserved, and for the FK dressed state to be IR finite. One can take (\ref{appendixeq1}) and Fourier transform to obtain 
\begin{equation}
    \begin{aligned}
        &[Q^+_{\varepsilon},\a_{1,\cI^+}(p)] = -\sqrt{2}\pi^2(1+|z_p|^2)\delta(\w_p-\w_0)\partial_{\bar{z}}\varepsilon(z_p,\bar{z}_p) \ ,\\
        &[Q^+_{\varepsilon},\a_{2,\cI^+}(p)] = -\sqrt{2}\pi^2(1+|z_p|^2)\delta(\w_p-\w_0)\partial_{z}\varepsilon(z_p,\bar{z}_p) \ ,\\
        &[Q^+_{\varepsilon},\a_{1,\cI^+}^{\dagger}(p)] = \sqrt{2}\pi^2(1+|z_p|^2)\delta(\w_p-\w_0)\partial_{z}\varepsilon(z_p,\bar{z}_p) \ ,\\
        &[Q^+_{\varepsilon},\a_{2,\cI^+}^{\dagger}(p)] = \sqrt{2}\pi^2(1+|z_p|^2)\delta(\w_p-\w_0)\partial_{\bar{z}}\varepsilon(z_p,\bar{z}_p) \ ,\\
    \end{aligned}
\end{equation}
It is these commutation relations that one can use to derive (\ref{appendixeq2}). Thus our framework 
is consistent with using the standard commutation relations in the bulk. If one instead uses 
\begin{equation}
    \begin{aligned}\label{lgt}
        &[Q^+_{\e}, A_z(u,y)]=-\partial_{z}\e(z_y,\bar{z}_y)\ ,\\
        &[Q^+_{\e}, A_{\bar{z}}(u,y)]=-\partial_{\bar{z}}\e(z_y,\bar{z}_y) \ ,\\
    \end{aligned}
\end{equation}
it is not clear how one can obtain (\ref{appendixeq2}).

Concretely, in the soft physics story~\cite{He_2014} the modified soft brackets are introduced to appropriately capture the large gauge transformation~\eqref{lgt} so that the transformation of the asymptotic states matches the one used to derive the charge from the covariant phase space formulation. This can be tracked due to the fact that soft modes of opposite helicities are not actually independent degrees of freedom~\cite{He_2014}. What we would like to point out here is that there still seems to be some open questions related to propagating these modified brackets to the way the Wilson line correlators are computed. In particular, as mentioned above, substituting the standard canonical commutation relations~\eqref{ccr} into the soft Wilson loop 
\begin{equation}\label{wline}
W= {\cal P}\exp\left[-ie\int_0^\infty ds \ n\cdot A(sn^\mu)e^{-\epsilon s}\right].
\end{equation} 
expressed as a path ordered exponential of the photon field~\eqref{aexp} along the trajectory of a charge moving with constant momentum matching the asymptotic state $n\propto p$ as in \cite{Hannesdottir:2019opa}, would seem to reproduce the expected phase rotation~\eqref{lgt}, in spite of~\eqref{appendixeq1} and the fact that we'd expect $\delta W$ to transform like $\delta A$ at the endpoints.  

Namely, there are some subtleties with propagating the modified soft brackets used in \cite{He_2014, Gabai_2016, strominger2018lecturesinfraredstructuregravity, Kapec_2017, Arkani_Hamed_2021} to their counterparts in the soft cloud. 
Importantly the Ward identities for the leading soft theorems can still be shown to follow with the usual brackets, despite the factor of $1/2$ in large gauge shift. 
Note that in \cite{Choi_2018} the authors also note that the standard commutation relations are required to obtain conservation for dressed states in gravity, where a similar story holds in the obvious manner. We suspect that better understanding these subtleties will be closely related to the tension of BMS frame dependence pointed out in~\cite{Elkhidir:2024izo}. We leave this to future work.
\newpage

\bibliographystyle{JHEP}
\bibliography{main.bib}

\providecommand{\href}[2]{#2}\begingroup\raggedright\begin{thebibliography}{10}

\bibitem{Strominger_2014}
A.~Strominger, \emph{On bms invariance of gravitational scattering}, \href{https://doi.org/10.1007/jhep07(2014)152}{\emph{Journal of High Energy Physics} {\bfseries 2014} (2014) }.

\bibitem{He_2014}
T.~He, P.~Mitra, A.P.~Porfyriadis and A.~Strominger, \emph{New symmetries of massless qed}, \href{https://doi.org/10.1007/jhep10(2014)112}{\emph{Journal of High Energy Physics} {\bfseries 2014} (2014) }.

\bibitem{Campiglia_2015}
M.~Campiglia and A.~Laddha, \emph{Asymptotic symmetries of gravity and soft theorems for massive particles}, \href{https://doi.org/10.1007/jhep12(2015)094}{\emph{Journal of High Energy Physics} {\bfseries 2015} (2015) 1–25}.

\bibitem{He_2015}
T.~He, V.~Lysov, P.~Mitra and A.~Strominger, \emph{Bms supertranslations and weinberg’s soft graviton theorem}, \href{https://doi.org/10.1007/jhep05(2015)151}{\emph{Journal of High Energy Physics} {\bfseries 2015} (2015) }.

\bibitem{Campiglia:2015qka}
M.~Campiglia and A.~Laddha, \emph{{Asymptotic symmetries of QED and Weinberg{\textquoteright}s soft photon theorem}}, \href{https://doi.org/10.1007/JHEP07(2015)115}{\emph{JHEP} {\bfseries 07} (2015) 115} [\href{https://arxiv.org/abs/1505.05346}{{\ttfamily 1505.05346}}].

\bibitem{Hu:2023geb}
Y.~Hu and S.~Pasterski, \emph{{Detector operators for celestial symmetries}}, \href{https://doi.org/10.1007/JHEP12(2023)035}{\emph{JHEP} {\bfseries 12} (2023) 035} [\href{https://arxiv.org/abs/2307.16801}{{\ttfamily 2307.16801}}].

\bibitem{Gonzalez:2025ene}
H.A.~Gonz{\'a}lez and J.~Salzer, \emph{{Energy Detectors and Asymptotic Symmetries}},  \href{https://arxiv.org/abs/2510.27348}{{\ttfamily 2510.27348}}.

\bibitem{Moult:2025njc}
I.~Moult, S.A.~Narayanan and S.~Pasterski, \emph{{Memory Correlators and Ward Identities in the `in-in' Formalism}},  \href{https://arxiv.org/abs/2512.02825}{{\ttfamily 2512.02825}}.

\bibitem{Himwich:2025ekg}
E.~Himwich and M.~Pate, \emph{{Light-ray Operators and the ${\rm w}_{1+\infty}$ Algebra}},  \href{https://arxiv.org/abs/2512.18973}{{\ttfamily 2512.18973}}.

\bibitem{Sheta:2025oep}
A.~Sheta, A.~Strominger, A.~Tropper and H.~Wei, \emph{{Soft Algebras in AdS$_4$ from Light Ray Operators in CFT$_3$}},  \href{https://arxiv.org/abs/2601.00096}{{\ttfamily 2601.00096}}.

\bibitem{Strominger:2026yrh}
A.~Strominger and H.~Wei, \emph{{EVERY CFT$_3$ HAS AN $ \mathcal{L}_{\Lambda} w_{1+\infty}$ SYMMETRY}},  \href{https://arxiv.org/abs/2603.26459}{{\ttfamily 2603.26459}}.

\bibitem{Hirai_2019}
H.~Hirai and S.~Sugishita, \emph{Dressed states from gauge invariance}, \href{https://doi.org/10.1007/jhep06(2019)023}{\emph{Journal of High Energy Physics} {\bfseries 2019} (2019) }.

\bibitem{Hirai_2021}
H.~Hirai and S.~Sugishita, \emph{Ir finite s-matrix by gauge invariant dressed states}, \href{https://doi.org/10.1007/jhep02(2021)025}{\emph{Journal of High Energy Physics} {\bfseries 2021} (2021) }.

\bibitem{Hirai_2023}
H.~Hirai and S.~Sugishita, \emph{Dress code for infrared safe scattering in qed}, \href{https://doi.org/10.1093/ptep/ptad057}{\emph{Progress of Theoretical and Experimental Physics} {\bfseries 2023} (2023) }.

\bibitem{PhysRevD.6.3543}
N.~Christ, B.~Hasslacher and A.H.~Mueller, \emph{Light-cone behavior of perturbation theory}, \href{https://doi.org/10.1103/PhysRevD.6.3543}{\emph{Phys. Rev. D} {\bfseries 6} (1972) 3543}.

\bibitem{BRANDT1971541}
R.A.~Brandt and G.~Preparata, \emph{Operator product expansions near the light cone}, \href{https://doi.org/https://doi.org/10.1016/0550-3213(71)90265-3}{\emph{Nuclear Physics B} {\bfseries 27} (1971) 541}.

\bibitem{Sterman:1975xv}
G.F.~Sterman, \emph{{Jet Structure in e+ e- Annihilation with Massless Hadrons}}, .

\bibitem{PhysRevD.19.2018}
C.L.~Basham, L.S.~Brown, S.D.~Ellis and S.T.~Love, \emph{Energy correlations in electron-positron annihilation in quantum chromodynamics: Asymptotically free perturbation theory}, \href{https://doi.org/10.1103/PhysRevD.19.2018}{\emph{Phys. Rev. D} {\bfseries 19} (1979) 2018}.

\bibitem{Hofman_2008}
D.M.~Hofman and J.~Maldacena, \emph{Conformal collider physics: energy and charge correlations}, \href{https://doi.org/10.1088/1126-6708/2008/05/012}{\emph{Journal of High Energy Physics} {\bfseries 2008} (2008) 012–012}.

\bibitem{Kravchuk_2018}
P.~Kravchuk and D.~Simmons-Duffin, \emph{Light-ray operators in conformal field theory}, \href{https://doi.org/10.1007/jhep11(2018)102}{\emph{Journal of High Energy Physics} {\bfseries 2018} (2018) }.

\bibitem{kologlu2020lightrayopeconformalcolliders}
M.~Kologlu, P.~Kravchuk, D.~Simmons-Duffin and A.~Zhiboedov, \emph{The light-ray ope and conformal colliders},  2020.

\bibitem{Moult:2025nhu}
I.~Moult and H.X.~Zhu, \emph{{Energy Correlators: A Journey From Theory to Experiment}},  \href{https://arxiv.org/abs/2506.09119}{{\ttfamily 2506.09119}}.

\bibitem{Caron_Huot_2023}
S.~Caron-Huot, M.~Koloğlu, P.~Kravchuk, D.~Meltzer and D.~Simmons-Duffin, \emph{Detectors in weakly-coupled field theories}, \href{https://doi.org/10.1007/jhep04(2023)014}{\emph{Journal of High Energy Physics} {\bfseries 2023} (2023) }.

\bibitem{li2025reggetrajectoriesdetectorsdistributions}
Y.-Z.~Li and D.~Simmons-Duffin, \emph{Regge trajectories, detectors, and distributions in the critical ${\rm o}(n)$ model},  2025.

\bibitem{herrmann2024energycorrelatorsperturbativequantum}
E.~Herrmann, M.~Kologlu and I.~Moult, \emph{Energy correlators in perturbative quantum gravity},  2024.

\bibitem{Ruan:2026xyd}
H.~Ruan, Y.~Zheng and H.X.~Zhu, \emph{{Celestial Energy-Energy Correlation in Yang-Mills Theory and Gravity}},  \href{https://arxiv.org/abs/2601.21852}{{\ttfamily 2601.21852}}.

\bibitem{Gonzo:2020xza}
R.~Gonzo and A.~Pokraka, \emph{{Light-ray operators, detectors and gravitational event shapes}}, \href{https://doi.org/10.1007/JHEP05(2021)015}{\emph{JHEP} {\bfseries 05} (2021) 015} [\href{https://arxiv.org/abs/2012.01406}{{\ttfamily 2012.01406}}].

\bibitem{Chicherin:2025keq}
D.~Chicherin, G.P.~Korchemsky, E.~Sokatchev and A.~Zhiboedov, \emph{{Energy correlators in four-dimensional gravity}},  \href{https://arxiv.org/abs/2512.23791}{{\ttfamily 2512.23791}}.

\bibitem{Choi_2018}
S.~Choi, U.~Kol and R.~Akhoury, \emph{Asymptotic dynamics in perturbative quantum gravity and bms supertranslations}, \href{https://doi.org/10.1007/jhep01(2018)142}{\emph{Journal of High Energy Physics} {\bfseries 2018} (2018) }.

\bibitem{1987Natur.327..123B}
V.B.~{Braginsky} and K.S.~{Thorne}, \emph{{Gravitational-wave bursts with memory and experimental prospects}}, \href{https://doi.org/10.1038/327123a0}{\emph{nat} {\bfseries 327} (1987) 123}.

\bibitem{PhysRevD.45.520}
K.S.~Thorne, \emph{Gravitational-wave bursts with memory: The christodoulou effect}, \href{https://doi.org/10.1103/PhysRevD.45.520}{\emph{Phys. Rev. D} {\bfseries 45} (1992) 520}.

\bibitem{Bieri_2013}
L.~Bieri and D.~Garfinkle, \emph{An electromagnetic analogue of gravitational wave memory}, \href{https://doi.org/10.1088/0264-9381/30/19/195009}{\emph{Classical and Quantum Gravity} {\bfseries 30} (2013) 195009}.

\bibitem{Bieri_2014}
L.~Bieri and D.~Garfinkle, \emph{Perturbative and gauge invariant treatment of gravitational wave memory}, \href{https://doi.org/10.1103/physrevd.89.084039}{\emph{Physical Review D} {\bfseries 89} (2014) }.

\bibitem{Prabhu:2022zcr}
K.~Prabhu, G.~Satishchandran and R.M.~Wald, \emph{{Infrared finite scattering theory in quantum field theory and quantum gravity}}, \href{https://doi.org/10.1103/PhysRevD.106.066005}{\emph{Phys. Rev. D} {\bfseries 106} (2022) 066005} [\href{https://arxiv.org/abs/2203.14334}{{\ttfamily 2203.14334}}].

\bibitem{Wightman1965FIELDSAO}
A.S.~Wightman and L.~Garding, \emph{Fields as operator-valued distributions in relativistic quantum theory},  1965.

\bibitem{1955NCimS...1..205L}
H.~{Lehmann}, K.~{Symanzik} and W.~{Zimmermann}, \emph{{Zur Formulierung quantisierter Feldtheorien}}, \href{https://doi.org/10.1007/BF02731765}{\emph{Nuovo Cimento Serie} {\bfseries 1} (1955) 205}.

\bibitem{PhysRev.112.669}
R.~Haag, \emph{Quantum field theories with composite particles and asymptotic conditions}, \href{https://doi.org/10.1103/PhysRev.112.669}{\emph{Phys. Rev.} {\bfseries 112} (1958) 669}.

\bibitem{ruelle1962asymptotic}
D.~Ruelle, \emph{On the asymptotic condition in quantum field theory}, {\emph{Helvetica Physica Acta} {\bfseries 35} (1962) 147}.

\bibitem{Dobrev:1977qv}
V.K.~Dobrev, G.~Mack, V.B.~Petkova, S.G.~Petrova and I.T.~Todorov, \emph{Harmonic Analysis on the n-Dimensional Lorentz Group and Its Application to Conformal Quantum Field Theory}, vol.~63, Springer (1977), \href{https://doi.org/10.1007/BFb0009678}{10.1007/BFb0009678}.

\bibitem{Sveshnikov_1996}
N.~Sveshnikov and F.~Tkachov, \emph{Jets and quantum field theory}, \href{https://doi.org/10.1016/0370-2693(96)00558-8}{\emph{Physics Letters B} {\bfseries 382} (1996) 403–408}.

\bibitem{Tkachov_1997}
F.V.~Tkachov, \emph{Measuring multijet structure of hadronic energy flow or, what is a jet?}, \href{https://doi.org/10.1142/s0217751x97002899}{\emph{International Journal of Modern Physics A} {\bfseries 12} (1997) 5411–5529}.

\bibitem{Korchemsky_1999}
G.P.~Korchemsky and G.~Sterman, \emph{Power corrections to event shapes and factorization}, \href{https://doi.org/10.1016/s0550-3213(99)00308-9}{\emph{Nuclear Physics B} {\bfseries 555} (1999) 335–351}.

\bibitem{Henriksson:2023cnh}
J.~Henriksson, P.~Kravchuk and B.~Oertel, \emph{{Missing local operators, zeros, and twist-4 trajectories}}, \href{https://doi.org/10.1007/JHEP07(2024)248}{\emph{JHEP} {\bfseries 07} (2024) 248} [\href{https://arxiv.org/abs/2312.09283}{{\ttfamily 2312.09283}}].

\bibitem{Li:2025knf}
Y.-Z.~Li and D.~Simmons-Duffin, \emph{{Regge trajectories, detectors, and distributions in the critical O(N) model}}, \href{https://doi.org/10.1007/JHEP02(2026)149}{\emph{JHEP} {\bfseries 02} (2026) 149} [\href{https://arxiv.org/abs/2506.06419}{{\ttfamily 2506.06419}}].

\bibitem{strominger2018lecturesinfraredstructuregravity}
A.~Strominger, \emph{Lectures on the infrared structure of gravity and gauge theory},  2018.

\bibitem{Korchemsky_2022}
G.P.~Korchemsky, E.~Sokatchev and A.~Zhiboedov, \emph{Generalizing event shapes: in search of lost collider time}, \href{https://doi.org/10.1007/jhep08(2022)188}{\emph{Journal of High Energy Physics} {\bfseries 2022} (2022) }.

\bibitem{Kulish:1970ut}
P.P.~Kulish and L.D.~Faddeev, \emph{{Asymptotic conditions and infrared divergences in quantum electrodynamics}}, \href{https://doi.org/10.1007/BF01066485}{\emph{Theor. Math. Phys.} {\bfseries 4} (1970) 745}.

\bibitem{PhysRev.140.B1110}
V.~Chung, \emph{Infrared divergence in quantum electrodynamics}, \href{https://doi.org/10.1103/PhysRev.140.B1110}{\emph{Phys. Rev.} {\bfseries 140} (1965) B1110}.

\bibitem{Gabai_2016}
B.~Gabai and A.~Sever, \emph{Large gauge symmetries and asymptotic states in qed}, \href{https://doi.org/10.1007/jhep12(2016)095}{\emph{Journal of High Energy Physics} {\bfseries 2016} (2016) }.

\bibitem{Satishchandran_2019}
G.~Satishchandran and R.M.~Wald, \emph{Asymptotic behavior of massless fields and the memory effect}, \href{https://doi.org/10.1103/physrevd.99.084007}{\emph{Physical Review D} {\bfseries 99} (2019) }.

\bibitem{Pasterski:2015zua}
S.~Pasterski, \emph{{Asymptotic Symmetries and Electromagnetic Memory}}, \href{https://doi.org/10.1007/JHEP09(2017)154}{\emph{JHEP} {\bfseries 09} (2017) 154} [\href{https://arxiv.org/abs/1505.00716}{{\ttfamily 1505.00716}}].

\bibitem{1962RSPSA.269...21B}
H.~{Bondi}, M.G.J.~{van der Burg} and A.W.K.~{Metzner}, \emph{{Gravitational Waves in General Relativity. VII. Waves from Axi-Symmetric Isolated Systems}}, \href{https://doi.org/10.1098/rspa.1962.0161}{\emph{Proceedings of the Royal Society of London Series A} {\bfseries 269} (1962) 21}.

\bibitem{1962RSPSA.270..103S}
R.K.~{Sachs}, \emph{{Gravitational Waves in General Relativity. VIII. Waves in Asymptotically Flat Space-Time}}, \href{https://doi.org/10.1098/rspa.1962.0206}{\emph{Proceedings of the Royal Society of London Series A} {\bfseries 270} (1962) 103}.

\bibitem{Christodoulou1989-1990}
D.~Christodoulou and S.~Klainerman, \emph{The global nonlinear stability of the minkowski space}, {\emph{S{\'e}minaire {\'E}quations aux d{\'e}riv{\'e}es partielles (Polytechnique)} (1989) 1}.

\bibitem{Ware_2013}
J.~Ware, R.~Saotome and R.~Akhoury, \emph{Construction of an asymptotic s matrix for perturbative quantum gravity}, \href{https://doi.org/10.1007/jhep10(2013)159}{\emph{Journal of High Energy Physics} {\bfseries 2013} (2013) }.

\bibitem{1974AZh....51...30Z}
Y.B.~{Zel'dovich} and A.G.~{Polnarev}, \emph{{Radiation of gravitational waves by a cluster of superdense stars}}, {\emph{Astronomicheskii Zhurnal} {\bfseries 51} (1974) 30}.

\bibitem{Strominger:2014pwa}
A.~Strominger and A.~Zhiboedov, \emph{{Gravitational Memory, BMS Supertranslations and Soft Theorems}}, \href{https://doi.org/10.1007/JHEP01(2016)086}{\emph{JHEP} {\bfseries 01} (2016) 086} [\href{https://arxiv.org/abs/1411.5745}{{\ttfamily 1411.5745}}].

\bibitem{Kapec_2017}
D.~Kapec, M.~Perry, A.-M.~Raclariu and A.~Strominger, \emph{Infrared divergences in qed revisited}, \href{https://doi.org/10.1103/physrevd.96.085002}{\emph{Physical Review D} {\bfseries 96} (2017) }.

\bibitem{Arkani_Hamed_2021}
N.~Arkani-Hamed, M.~Pate, A.-M.~Raclariu and A.~Strominger, \emph{Celestial amplitudes from uv to ir}, \href{https://doi.org/10.1007/jhep08(2021)062}{\emph{Journal of High Energy Physics} {\bfseries 2021} (2021) }.

\bibitem{Hannesdottir:2019opa}
H.~Hannesdottir and M.D.~Schwartz, \emph{{$S$ -Matrix for massless particles}}, \href{https://doi.org/10.1103/PhysRevD.101.105001}{\emph{Phys. Rev. D} {\bfseries 101} (2020) 105001} [\href{https://arxiv.org/abs/1911.06821}{{\ttfamily 1911.06821}}].

\bibitem{Elkhidir:2024izo}
A.~Elkhidir, D.~O'Connell and R.~Roiban, \emph{{Supertranslations from Scattering Amplitudes}},  \href{https://arxiv.org/abs/2408.15961}{{\ttfamily 2408.15961}}.

\end{thebibliography}\endgroup

\end{document}